\documentclass[preprint,nocopyrightspace,numbers,10pt]{sigplanconf-pldi17}

\usepackage{courier}            % standard fixed width font
\usepackage[scaled]{helvet} % see www.ctan.org/get/macros/latex/required/psnfss/psnfss2e.pdf
\usepackage[hyphens]{url}                  % format URLs
\usepackage{listings}          % format code
\usepackage{enumitem}      % adjust spacing in enums
\usepackage{algorithm,algorithmicx}
\usepackage{amsthm}
\usepackage{graphicx}
\usepackage{comment}
\usepackage{amsmath}
\usepackage{hyperref}
\usepackage{bbm, dsfont}
\usepackage{mathrsfs}
\usepackage{xcolor}
\usepackage{esvect}
\usepackage{xspace}
\usepackage{array, multirow} % multirow in table
\usepackage{rotating, makecell} % cell rotation
\usepackage[noend]{algpseudocode}
\usepackage{balance}
\usepackage{qtree}
\usepackage{amssymb}
\usepackage{multirow}
\usepackage{fancyvrb}
\usepackage{tikz}
\usetikzlibrary{positioning,shapes,arrows}
\usepackage{pgfplots}

\newcommand{\comps}{\Lambda}
\newcommand{\prog}{\mathcal{P}}

\newcommand{\sketch}{\mathcal{S}}
\newcommand{\project}{\pi}
\newcommand{\select}{\sigma}

\newcommand{\constraint}{\phi}
\newcommand{\comp}{\mathcal{X}}

\newcommand{\abs}{\alpha}
\newcommand{\inhabit}{\Omega}
\newcommand{\tab}{\textsf{T}}

\newcommand{\sample}{\mathcal{E}}
\newcommand{\hypo}{\mathcal{H}}
\newcommand{\var}{\upsilon}
\newcommand{\subprog}{\rho}
\newcommand{\qual}{\mathcal{Q}}

\newcommand{\cols}{\kappa}
\newcommand{\contents}{\varsigma}
\newcommand{\ex}{\sample}
\newcommand{\rows}{\small \textsf{row}}
\newcommand{\columns}{\small \textsf{col}}
\newcommand{\groups}{\small \textsf{group}}
\newcommand{\cells}{\small \textsf{newVals}}
\newcommand{\heads}{\small \textsf{newCols}}
\newcommand{\eval}[1]{{[\![#1]\!]}}
\newcommand{\peval}[1]{{[\![#1]\!]_\partial}}
\newcommand{\worklist}{W}
\newcommand{\toConstraint}{\Phi}
\newcommand{\complete}{\mathcal{C}}

\usepackage{pifont}% http://ctan.org/pkg/pifont
\newcommand{\cmark}{\ding{51}}%
\newcommand{\xmark}{\ding{55}}%

\newcommand*{\comments}{} %comment to remove comments
\ifdefined\comments
\newcommand{\yu}[1]{\textcolor{blue}{\textbf{YU:} #1}}
\newcommand{\ruben}[1]{\textcolor{magenta}{\textbf{RUBEN:} #1}}
\newcommand{\isil}[1]{\textcolor{red}{\textbf{I\c{S}IL:} #1}}
\newcommand{\todo}[1]{\textcolor{red}{#1}}
\else
\newcommand{\yu}[1]{}
\newcommand{\ruben}[1]{}
\newcommand{\isil}[1]{}
\newcommand{\todo}[1]{}
\fi

\newtheorem{example}{Example}
\newtheorem{definition}{Definition}

\newcommand{\toolname}{{\sc Morpheus}\xspace}
\newcommand{\sql}{{\sc SQLSynthesizer}\xspace}

\usepackage{xcolor,colortbl}
\definecolor{babypurple}{RGB}{222,201,255}
\definecolor{lightgoldenrodyellow}{rgb}{0.98, 0.98, 0.82}
\definecolor{babyred}{RGB}{220,210,214}
\definecolor{camouflagegreen}{rgb}{0.47, 0.53, 0.42}
\definecolor{babygreen}{rgb}{0.47, 0.53, 0.42}
\definecolor{camel}{rgb}{0.76, 0.6, 0.42}
\definecolor{darkpastelpurple}{rgb}{0.59, 0.44, 0.84}

\algrenewcommand{\algorithmiccomment}[1]{\hfill $\triangleright${#1}}

\begin{document}

\authorinfo{Yu Feng}{University of Texas at Austin, USA}{yufeng@cs.utexas.edu}
\authorinfo{Ruben Martins}{University of Texas at Austin, USA}{rmartins@cs.utexas.edu}
\authorinfo{Jacob Van Geffen}{University of Texas at Austin, USA}{jsv@cs.utexas.edu}
\authorinfo{Isil Dillig}{University of Texas at Austin, USA}{isil@cs.utexas.edu}
\authorinfo{ Swarat Chaudhuri}{Rice University, USA}{swarat@rice.edu}

%
% any author declaration will be ignored  when using 'plid' option (for double blind review)
%

%\title{Automating Data Preparation Tasks from Input-Output Examples}

\title{Component-based Synthesis of Table Consolidation and Transformation Tasks from Examples
%\vspace{-0.4in}
}

\maketitle
\begin{abstract}
This paper presents an example-driven synthesis technique for automating a large class of data preparation tasks that arise in data science. Given a set of input tables  and an output table, our approach synthesizes a table transformation program that performs the desired task. Our approach is not restricted to a fixed set of DSL constructs and can synthesize programs from an arbitrary set of components, including higher-order combinators. At a high-level, our approach performs type-directed enumerative search over partial programs but incorporates two key innovations that allow it to scale: First, our technique can utilize any first-order specification of the components and uses SMT-based deduction to reject partial programs. Second, our algorithm uses partial evaluation to increase the power of deduction and drive enumerative search. We have evaluated our  synthesis algorithm on dozens of data preparation tasks obtained from on-line forums, and we show that our approach can automatically solve  a large class of  problems encountered by R users.
\end{abstract}

\newcommand{\irule}[2]%
   {\mkern-2mu\displaystyle\frac{#1}{\vphantom{,}#2}\mkern-2mu} 
\newcommand{\irulelabel}[3]
{
\mkern-2mu
\begin{array}{ll}
\displaystyle\frac{#1}{\vphantom{,}#2} & #3
\end{array} 
\mkern-2mu
} 

\section{Introduction}\label{sec:intro}

\begin{figure*}[t]
\begin{center}
\includegraphics[scale=0.35]{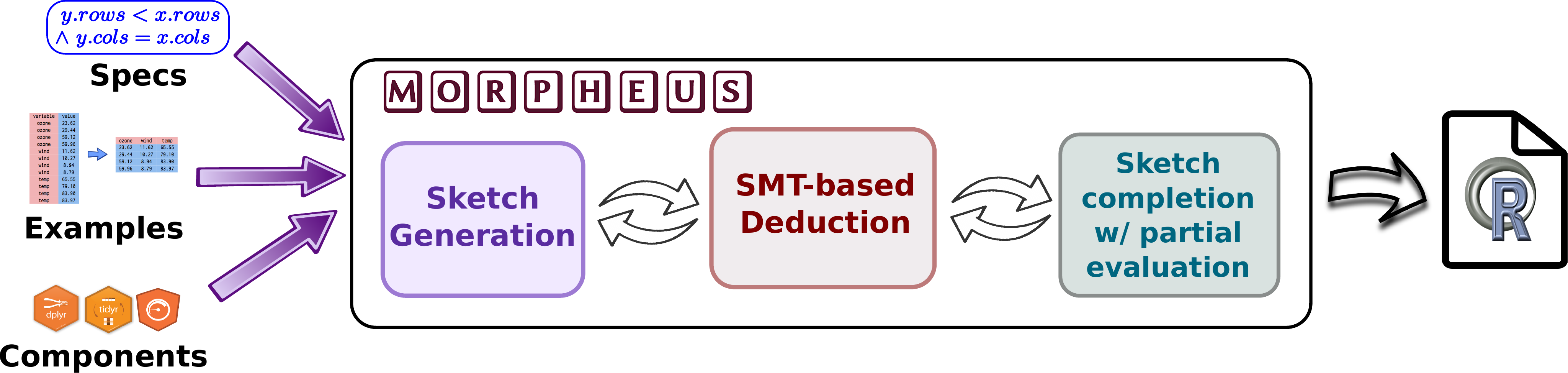}
\end{center}
\vspace{-0.1in}
\caption{Overview of our approach}\label{fig:overview}
\vspace{-0.1in}
\end{figure*}

Due to the explosion in the amount of available data over the last decade,  \emph{data analytics} has gained enormous popularity across a wide range of industries. While the ultimate goal of data analytics is to discover hidden patterns in existing data and perform predictive modeling through machine learning,  an important precursor to these tasks is \emph{data preparation}. Generally speaking, many data preparation tasks involve consolidating multiple data sources into a single table (called \emph{data frame} in the popular R language), reshaping data from one format into another, or adding new rows or columns to an existing table. Even though data preparation is considered ``janitor work" of data science, most data scientists spend over 80\% of their time in converting raw data into a form that is suitable for an analysis or visualization task~\cite{dasu}.

\begin{comment}
Due to the explosion in the amount of available data over the last decade,  \emph{data analytics} has gained enormous popularity across a wide range of industries, ranging from healthcare to small businesses. In particular, data analytics makes it possible to discover hidden patterns in existing data and perform predictive modeling through the use of machine learning algorithms. Because of this rising demand for data analytics, domain-specific languages like R have gained popularity among users hoping to gain insights from their data.

While DSLs like R provide a large collection of data mining and machine learning algorithms, an important precursor to data analytics is \emph{data preparation}. For instance, since the original raw data is often in a format that is not immediately suitable for analysis, data scientists must massage the data into a form that is considered ``tidy", where columns correspond to variables and rows represent observations. Generally speaking, many data preparation tasks involve consolidating multiple data sources into a single table (called \emph{data frame} in R), reshaping data from one format into another, or adding new rows or columns to an existing table. Even though data preparation is considered ``janitor work" of data science, most data scientists spend over 80\% of their time in converting raw data into a form that is suitable for an analysis or visualization task~\cite{dasu}.
\end{comment}

In this paper, we propose a novel program synthesis technique for automating a large class of data preparation tasks. Given a set of input tables (or data frames) and a desired output table, our technique  synthesizes a table transformation program that automates the desired task. While there has been some previous work on automated synthesis of table transformations from input-output examples (e.g., ~\cite{example-pldi11,zhang-ase2013}), existing techniques focus on narrowly-defined domain-specific languages (DSLs), such as subsets of the Excel macro language~\cite{example-pldi11} or fragments of SQL~\cite{zhang-ase2013}. Unfortunately, many common data preparation tasks (e.g., those that involve reshaping tables or require performing nested table joins) fall outside the scope of these previous approaches.

In order to support a large class of data preparation tasks, we propose a flexible \emph{component-based} approach to synthesize programs that operate over tables. In contrast to previous techniques, our method does not assume a fixed DSL and is parametrized over a set of components, which can be extended over time as new libraries emerge or customized by  users. Furthermore, these components can include both higher-order and first-order combinators.
 As we demonstrate empirically, the generality of our technique allows the automation of a diverse class of data preparation tasks involving data tidying, reshaping, consolidation, and computation.

 While our more general formulation of the problem increases the applicability of the approach,  it also brings a number of algorithmic challenges. First, our synthesis algorithm cannot exploit ``hard-coded" assumptions about specific components or DSL constructs. Second, in order to have a chance of scaling, the algorithm must be able to utilize arbitrary first-order specifications of the components.
 
%$ Second, the synthesis algorithm should be  scalable and expressive enough to support the kinds of components that are provided by modern data preparation libraries.

 Our synthesis approach solves these challenges through a number of algorithmic innovations.  %At a high level, our algorithm follows a strategy of
Similar to multiple recent approaches to synthesis~\cite{example-cav13,lambda2,myth}, our technique performs type-directed enumerative search through a space of {\em partial programs}.  However, one of key technical innovations underlying our approach is to use  \emph{SMT-based deduction to reject
partial programs}.  While previous synthesis techniques (e.g.,
\cite{lambda2,flashmeta}) have used deduction to speed up search,
their deductive reasoning capabilities are hard-wired to a fixed set
of DSL constructs.  Consequently, adding new components 
to the DSL require changing the underlying synthesis algorithm. In
contrast, our approach can apply deductive reasoning to any component
that is equipped with a corresponding first-order specification,
%For instance, we have written specifications of methods
%provided by two popular R libraries for data tidying and manipulation,
%and we show that our synthesis algorithm can successfully synthesize a
and is able to synthesize a large class of table transformation programs despite the lack of
\emph{any} hard-coded component-specific reasoning.

The second key insight underlying our technique is decompose the synthesis task into two separate \emph{sketch generation} and \emph{sketch completion} phases, with the goal of achieving better scalability. Specifically, we observe that the components  used in data preparation tasks can be classified into two classes, namely \emph{table transformers}  and \emph{value transformers}. Table transformers (e.g., select and join from relational algebra) are higher-order functions that change the shape of the input tables, whereas value transformers (e.g., {\tt MAX}, {\tt MEAN}) are first-order operators that are supplied as arguments to the table transformers.  Our synthesis algorithm first generates program sketches that fully specify the set of table transformers used in the program, and the subsequent sketch completion phase instantiates the holes with programs constructed using the first-order value transformers. The key advantage of this decomposition is that we can reject program sketches using SMT-based deduction. Since a sketch can be completed in \emph{many} possible ways, SMT-based refutation of a sketch allows us to dramatically prune the search space.

The third crucial ingredient of our synthesis algorithm is the use of \emph{partial evaluation} to complete program sketches. Given a partially-filled program sketch, our approach uses the input tables provided by the user to evaluate subterms of the sketch. In this context, the use of partial evaluation has two key benefits: First, because partial evaluation allows us to obtain \emph{concrete tables} for subterms in the sketch, we can perform more precise deductive reasoning, which allows us to refute partially filled sketches that we could not reject otherwise. Second, partial evaluation is used to drive enumerative search by finitizing the universe of constants from which expressions are constructed. For instance, the use of partial evaluation allows us to determine the universe of strings from which column names can be selected.

%To illustrate this point, consider the statement {\tt select}($t$, $\phi$) where $t$ is a term of type table and $\phi$ is a predicate of the form ${\tt column} > {\tt int}$. Since there are infinitely many integers that can appear in $\phi$, our synthesis algorithm would either have to enumerate all integers or ask the user to provide the set of  constants that are allowed to appear in the program.  Because partial evaluation allows us to concretize term $t$ to a table {\tt T}, it suffices to consider only those integers that appear as values in {\tt T}. In particular, if $c'$ is a constant that does not appear in {\tt T}, the predicate ${\tt column} > c'$ is \emph{ equivalent modulo the inputs} to another predicate ${\tt column} > c$, where $c$ is a constant that does appear in {\tt T}.  Hence, the use of partial evaluation allows us to finitize the universe of constants without losing completeness with the respect to the input-output examples.

Figure~\ref{fig:overview} shows a schematic overview of our approach, implemented in a tool called \toolname. The  input to \toolname is a set of input tables together with the desired output table. Additionally, the user can also provide a set of components (i.e., library methods), optionally with their corresponding first-order specification. Since our implementation already comes with a built-in set of components that are commonly used in data preparation, the user does not need to provide any additional components but can do so if she  so desires. As shown in Figure~\ref{fig:overview}, our approach decomposes the underlying synthesis task into two separate sketch generation and sketch completion phases, both of which utilize SMT-based deduction to refute partial programs.

We have evaluated our approach on a suite of data preparation tasks
for the R programming language, drawn from discussions among R users
in on-line forums such as Stackoverflow.  The ``components'' in our
evaluation are methods provided by two popular R libraries, namely
{\tt tidyr} and {\tt dplyr}, for data tidying and manipulation.  Our
experiments show that \toolname \ can successfully synthesize a diverse
class of real-world data preparation programs. We also demonstrate
that SMT-based deduction and partial evaluation are crucial for the
scalability of our approach.

This paper makes the following  key contributions:

\begin{itemize}
\item We propose a programming-by-example methodology for automating table
  transformation and consolidation tasks that commonly arise in  data
  preparation. 
\item We describe a novel component-based synthesis algorithm that
  uses SMT-based deduction and partial evaluation to dramatically
  prune the search space.
\item We implement these ideas in a tool called \toolname 
and demonstrate that our approach can be
  used to synthesize a wide variety of data preparation tasks in R.
  %using
  %the components provided by the {\tt tidyr} and {\tt dplyr}
  %libraries.
\end{itemize}

\section{Motivating Examples}
\label{sec:example}

%Performing data preparation tasks is often beyond the knowledge of most R users and there are thousands of posts on Stackoverflow with questions on how to perform complex data transformation tasks. In this section, we illustrate the synthesis capabilities of \toolname using three examples from Stackoverflow.

In this section, we illustrate the diversity of data preparation tasks using a few  examples collected from Stackoverflow.

\begin{example}\label{ex:long-to-wide}
An R user has the  data frame in Figure~\ref{fig:ex}(a),  but wants to transform it to  the following format~\cite{ex1}:

{\small 
    \begin{center}
   % \begin{flushleft}{\hspace{0.4in}Desired data:}\end{flushleft}
   % \vspace{0.1in}
    \begin{tabular}{c c c c c} 
     \hline
     \rowcolor{lightgoldenrodyellow}
     id & A\_2007 & B\_2007 & A\_2009 & B\_2009 \\ [0.5ex] 
     \hline\hline
     \rowcolor{camouflagegreen}
     1 & 5 & 10 & 5 & 17\\
     \hline
     \rowcolor{babypurple}
     2 & 3 & 50 & 6 & 17\\
     \hline
    \end{tabular}
    \end{center}
    }
 
 \noindent
 Even though the user is  quite familiar with R  libraries for data preparation, she is still not able to perform the desired task. Given this example, \toolname can automatically synthesize the following R program:

%\toolname can help thousands of R users like Bob by automatically synthesizing R programs that will perform the desired data transformation. To use \toolname, Bob only needs to provide input-output examples. In this case, consider that Bob has given \toolname the input-output examples presented above. \

%\scalebox{0.9}
{
\small
\begin{Verbatim}[commandchars=\\\{\},codes={\catcode`$=3\catcode`_=8}]
df1=\textbf{gather}(input,var,val,id,A,B)
df2=\textbf{unite}(df1,yearvar,var,year)
df3=\textbf{spread}(df2,yearvar,val)
\end{Verbatim}
}

Observe that this example requires both reshaping the table and appending contents of some cells to column names.
\end{example}

%The user can now use 

%Observe that writing this code is non-trivial for a user like Bob since it involves complex data frame transformations such as {\tt gather} and {\tt spread}. Gather takes multiple columns and collapses them into key-value pairs, duplicating all other columns as needed. After using {\tt gather}, Bob would need to use {\tt unite} to merge the contents of two columns into one new column. Finally, he would need to use the {\tt spread} operation to spread a key-value pair across multiple columns and get the desired data frame.

\begin{example}\label{ex:flights}Another R user has the data frame from Figure~\ref{fig:ex}(b) and wants to compute, for each source location $L$, the number and percentage of flights that go to Seattle (SEA)  from $L$~\cite{ex2}.
 In particular, the  output should be as follows:

{\small 
\begin{center}
\begin{tabular}{c c c} 
     \hline
     \rowcolor{lightgoldenrodyellow}
     origin & n & prop \\ [0.5ex] 
     \hline\hline
     \rowcolor{babygreen}
     EWR   & 2  & 0.6666667\\
     \hline
    \rowcolor{babypurple}
     JFK   & 1  & 0.3333333\\
     \hline
    \end{tabular}
    \end{center}
}

\toolname can automatically synthesize the following R program to extract the desired information:

%df2=\textbf{group\_by}(origin)
{
\small
\begin{Verbatim}[commandchars=\\\{\},codes={\catcode`$=3\catcode`_=8}]
df1=\textbf{filter}(input, dest \textbf{==} "SEA")
df2=\textbf{summarize}(\textbf{group\_by}(df1, origin), n = \textbf{n()})
df3=\textbf{mutate}(df2, prop = n \textbf{/} \textbf{sum}(n))
\end{Verbatim}
}

Observe that this example involves selecting a subset of the data and performing some computation on that subset.

\end{example}

\begin{example} A data analyst has the following raw data about the position of vehicles for a driving simulator~\cite{ex3}:

\newcommand{\mc}[2]{\multicolumn{#1}{c}{#2}}
\newcolumntype{a}{>{\columncolor{babyred}}c}
\newcolumntype{b}{>{\columncolor{babypurple}}c}
\newcommand{\done}{\cellcolor{babyred}done}  %{0.9}

\vspace{0.1in}
\begin{tabular}{ll}
\hspace{-0.05in}Table 1: & \hspace{.95in}Table 2:
\end{tabular}
\vspace{-0.1in}

{\small 
\begin{center}
\begin{tabular}{ll}
%\begin{flushleft}{\hspace{0.4in}Table 1:}\end{flushleft}
%\vspace{0.1in}
\begin{tabular}{a c c c} 
     \hline
     \cellcolor{lightgoldenrodyellow}{frame} & \cellcolor{babyred}{X1} & \cellcolor{babyred}{X2} & \cellcolor{babyred}{X3} \\ [0.5ex]
     \hline\hline
     %\rowcolor{babygreen}
     \cellcolor{lightgoldenrodyellow}{1} & 0 & 0 & 0\\
     \hline
     %\rowcolor{babypurple}
     \cellcolor{lightgoldenrodyellow}{2} & \cellcolor{babypurple}{10} & \cellcolor{babygreen}{15} & 0\\
     \hline
     %\rowcolor{babyred}
     \cellcolor{lightgoldenrodyellow}{3} & \cellcolor{babygreen}{15} & \cellcolor{babypurple}{10} & 0\\
     \hline
    % \cellcolor{lightgoldenrodyellow}{4} & \cellcolor{babygreen}{15} & \cellcolor{babypurple}{10} & 0\\
    % \hline
    \end{tabular}

    \hspace*{0.01in}
    \begin{tabular}{c c c c} 
     \hline
     %\rowcolor{lightgoldenrodyellow}
     \cellcolor{lightgoldenrodyellow}{frame} & \cellcolor{babyred}{X1} & \cellcolor{babyred}{X2} & \cellcolor{babyred}{X3} \\ [0.5ex] 
     \hline\hline
     %\rowcolor{babygreen}
     \cellcolor{lightgoldenrodyellow}{1} & 0 & 0 & 0\\
     \hline
     %\rowcolor{babypurple}
     \cellcolor{lightgoldenrodyellow}{2} & \cellcolor{babypurple}{14.53} & \cellcolor{babygreen}{12.57} & 0\\
     \hline
     %\rowcolor{babyred}
     \cellcolor{lightgoldenrodyellow}{3} & \cellcolor{babygreen}{13.90} & \cellcolor{babypurple}{14.65} & 0\\
     \hline
     %\cellcolor{lightgoldenrodyellow}{4} & \cellcolor{babygreen}{14.10} & \cellcolor{babypurple}{14.70} & 0\\
     %\hline
    \end{tabular}
    
    \end{tabular}
    \end{center}
    }

Here, Table 1 contains the unique identification number for each vehicle (e.g., 10, 15), with 0 indicating the absence of a vehicle.
The column labeled ``frame'' in Table 1 measures the time step, and the columns ``X1'', ``X2'', ``X3'' track which vehicle is closer to the driver. For example, at frame 3, the vehicle with ID 15 is the closest to the driver. Table 2 has a similar structure as Table 1 but contains the speeds of the vehicles instead of their identification number. For example, at frame 3, the speed of the vehicle with ID 15 is 13.90 m/s.
The data analyst wants to consolidate these two data frames into a new table with the following shape:

{\small 
    \begin{center}
   % \begin{flushleft}{\hspace{0.85in}Desired data:}\end{flushleft}
   % \vspace{0.1in}
    \begin{tabular}{c c c c} 
     \hline
     \rowcolor{lightgoldenrodyellow}
     frame & pos & carid & speed \\ [0.5ex] 
     \hline\hline
     \rowcolor{babygreen}
     \cellcolor{lightgoldenrodyellow}{2} & \cellcolor{babyred}{X1} & \cellcolor{babypurple}{10} & \cellcolor{babypurple}{14.53}\\
     \hline
     \rowcolor{babygreen}
     \cellcolor{lightgoldenrodyellow}{3} & \cellcolor{babyred}{X2} & \cellcolor{babypurple}{10} & \cellcolor{babypurple}{14.65}\\
     \hline
     \rowcolor{babygreen}
     %\cellcolor{lightgoldenrodyellow}{4} & \cellcolor{babyred}{X2} & \cellcolor{babypurple}{10} & \cellcolor{babypurple}{14.70}\\
     \hline
     \rowcolor{babygreen}
     \cellcolor{lightgoldenrodyellow}{2} & \cellcolor{babyred}{X2} & \cellcolor{babygreen}{15} & \cellcolor{babygreen}{12.57}\\
     \hline
     \rowcolor{babygreen}
     \cellcolor{lightgoldenrodyellow}{3} & \cellcolor{babyred}{X1} & \cellcolor{babygreen}{15} & \cellcolor{babygreen}{13.90}\\
     \hline
     \rowcolor{babygreen}
     %\cellcolor{lightgoldenrodyellow}{4} & \cellcolor{babyred}{X1} & \cellcolor{babygreen}{15} & \cellcolor{babygreen}{14.10}\\
     \hline
    \end{tabular}
    \end{center}
    }

Despite looking into  R libraries for data preparation, the  analyst still cannot figure out how to perform this task and asks for help on Stackoverflow. \toolname can  synthesize the following R program to automate this complex task:
%~\footnote{\url{http://stackoverflow.com/questions/32875699/how-to-combine-two-data-frames-in-r-see-details}} \toolname can solve this problem in \todo{xxx} seconds with the following R program:

{
\small
\begin{Verbatim}[commandchars=\\\{\},codes={\catcode`$=3\catcode`_=8}]
df1=\textbf{gather}(table1,pos,carid,X1,X2,X3)
df2=\textbf{gather}(table2,pos,speed,X1,X2,X3)
df3=\textbf{inner\_join}(df1,df2)
df4=\textbf{filter}(df3,carid \textbf{!=} 0)
df5=\textbf{arrange}(df4,carid,frame)
\end{Verbatim}
}

%This complex data transformation combines the reshaping operator {\tt gather} and the consolidation operator {\tt inner} {\tt\_join} with the {\tt filter} and {\tt arrange} operators. The {\tt filter} operator selects a subset of the rows of a data frame. Since Tom only wants rows that have ``carid'' different than 0, \toolname must remove those rows from the desired data frame.
%
%The {\tt arrange} operator reorders the rows of a data frame. In this example, Tom wants the final data frame to be ordered by ascending order of the ``carid'' and ``frame'' values. Finding such large sequence of operators that combine different aspects of data preparation can be quite challenging for most R users. By automatically synthesizing R programs from input-output examples, \toolname will significantly increase productivity of thousands of R users on their daily data preparation tasks.
\end{example}

\begin{figure}
\footnotesize
\begin{tabular}{ c  c }
%(a) & (b) \\
\begin{minipage}{0.4\linewidth}
\hspace*{-10pt}
{\small 
\begin{center}
    %\begin{flushleft}{\hspace{1.1in}Raw data:}\end{flushleft}
    %\vspace{0.1in}
\begin{tabular}{c c c c} 
     %Raw data:\\ [0.5ex]
     \hline
     \rowcolor{lightgoldenrodyellow}
     id & year & A & B \\ [0.5ex] 
     \hline\hline
     \rowcolor{babygreen}
     1 & 2007 & 5 & 10\\
     \hline
     \rowcolor{babypurple}
     2 & 2009 & 3 & 50\\
     \hline
     \rowcolor{babygreen}
     1 & 2007 & 5 & 17\\
     \hline
     \rowcolor{babypurple}
     2 & 2009 & 6 & 17\\
     \hline
    \end{tabular}
    \end{center}
    }
\end{minipage}
&
\begin{minipage}{0.5\linewidth}
\hspace*{-5pt}
{\small 
\begin{center}
\begin{tabular}{c c c} 
     \hline
     \rowcolor{lightgoldenrodyellow}
     flight & origin & dest \\ [0.5ex] 
     \hline\hline
     \rowcolor{babygreen}    
     11   & EWR  & SEA\\
     \hline
      \rowcolor{white}
     % 1141 &  JFK  & MIA\\
      725 &  JFK  & BQN\\
      \hline
      \rowcolor{babypurple}
    495   & JFK  & SEA\\
    \hline
    \rowcolor{white}
     461  &  LGA  & ATL\\
     \hline
    1696  &  EWR  & ORD\\
    \hline
    \rowcolor{babygreen}
   1670   & EWR  & SEA\\
     \hline
    \end{tabular}
    \end{center}
}
\end{minipage}
\\
(a) & (b) 
\end{tabular}
\caption{(a) Data frame for Example~\ref{ex:long-to-wide}; (b) for Example~\ref{ex:flights}.}
\label{fig:ex}
\vspace{-5pt}
\end{figure}

\section{Problem Formulation}\label{sec:problem}
In order to precisely describe our synthesis problem, we first present some definitions that we  use throughout the paper.

\begin{definition}{{(\bf Table)}}
A \emph{table} $\tab$ is a tuple $(r, c, \tau, \contents)$ where:
\begin{itemize}
\item $r, c$ denote  number of rows and columns  respectively
\item $\tau: \{l_1: \tau_i, \ldots, l_n: \tau_n\}$ denotes the type of $\tab$. In particular, each $l_i$ is the name of a column in $\tab$ and $\tau_i$ denotes the type of the value stored in $\tab$. We assume that each $\tau_i$ is either {\tt num} or {\tt string}.
\item $\contents$ is a mapping from each cell $(i, j) \in ([0, r) \times [0, c))$ to a value $v$ stored in that cell
\end{itemize}
\end{definition}
Given a table $\tab = (r, c, \tau, \contents)$, we write $\tab.\rows$ and $\tab.\columns$ to denote $r$ and $c$ respectively. We also write $\tab_{i,j}$ as shorthand for $\contents(i, j)$ and $\emph{type}(\tab)$ to represent $\tau$. We  refer to all record types  $\{l_1: \tau_i, \ldots, l_n: \tau_n\}$ as type {\tt tbl}.
In addition, tables with only one row are referred to as being of type {\tt row}.

\begin{definition}({\bf Component})
A component $\comp$ is a triple $(f, \tau, \phi)$ where $f$ is a string denoting $\comp$'s name, $\tau$ is the type signature (see Figure~\ref{fig:types}), and $\phi$ is a first-order formula that specifies $\comp$'s input-output behavior.
\end{definition}

Given a component $\comp = (f, \tau, \phi)$, the specification $\phi$ is over the vocabulary $x_1, \ldots, x_n, y$, where $x_i$ denotes $\comp$'s $i$'th argument and $y$ denotes $\comp$'s return value. Note that  specification $\phi$ does not need to \emph{precisely} capture $\comp$'s input-output behavior; it only needs to be an \emph{overapproximation}. Thus, $\emph{true}$ is always a valid specification for any component. 

With slight abuse of notation, we sometimes write $\comp(\ldots)$ to mean $f(\ldots)$ whenever $\comp = (f, \tau, \phi)$.  Also, given a component $\comp$ and arguments $c_1, \ldots, c_n$, we write $\eval{\comp(c_1, \ldots, c_n)}$  to denote the result of evaluating $\comp$ on arguments $c_1, \ldots, c_n$.

\begin{definition}{{(\bf Problem specification)}}
The \emph{specification} for a synthesis problem is a pair $(\ex, \comps)$ where:
\begin{itemize}
\item $\ex$ is an input-output example $(\vec{\tab}_\emph{in}, \tab_\emph{out})$ such that $\vec{\tab}_\emph{in}$ denotes a list of input tables, and  $\tab_\emph{out}$ is the output table,
\item $\comps = (\comps_\tab \cup \comps_v)$ is a set of components, where $\comps_\tab, \comps_v$ denote \emph{table transformers} and  \emph{value transformers} respectively. We assume that $\comps_\tab$ includes higher-order functions, but $\comps_v$ consists of first-order operators.
\end{itemize}
\end{definition}

Given an input-output example $\ex = (\vec{\tab}_\emph{in}, \tab_\emph{out})$, we write $\ex_{in}$, $\ex_{out}$ to denote 
$\vec{\tab}_\emph{in}$, $\tab_\emph{out}$ respectively.
%Also, we classify components $\comps$ into two disjoint classes $\comps_\tab$ and $\comps_v$, where $\comps_\tab$ denotes \emph{table transformer} components that take at least one table as an argument and return a table. Components of all other types are \emph{value transformers} $\comps_v$. 
As mentioned in Section~\ref{sec:intro}, we distinguish between the  higher-order \emph{table transformers} $\comps_\tab$ and first-order \emph{value transformers} $\comps_v$.   In the rest of the paper, we assume that table transformers $\comps_\tab$ only take tables and first-order functions (constructed using constants and components in $\comps_v$) as arguments.
%Since table transformers are typically higher-order functions, components in $\comps_v$ are  used as arguments for components in $\comps_\tab$. 
%We assume that components in $\comps_v$ are first-order functions and do not return objects of type {\tt table}. 

\begin{example}
Consider the  selection operator $\sigma$ from relational algebra, which takes a table and a predicate and returns a table. In our terminology, such a component is a higher-order \emph{table transformer}. In contrast, an aggregate function such as \emph{sum} that takes a list of values and returns their sum is a \emph{value transformer}. Similarly, the boolean operator $\geq$ is also a value transformer.
\end{example}

\begin{definition}{{(\bf Synthesis problem)}} Given specification $(\ex, \comps)$ where $\ex = (\vec{\tab}_\emph{in}, \tab_\emph{out})$, the synthesis problem is to infer a program $\lambda \vec{x}. e$ such that (a) $e$ is a well-typed expression over components in $\comps$, and
(b)  $(\lambda \vec{x}. e) \vec{\tab}_\emph{in} = \tab_\emph{out}$.
\end{definition}

\begin{comment}
The synthesis problems we consider in this paper take a \emph{single} input-output example rather than a \emph{set} of examples because all data preparation questions that we have encountered on on-line forums come with a single example. However, we do not restrict table sizes or the number of input tables. While it is straightforward to extend our technique to multiple examples, the single-example assumption  simplifies our technical presentation and better matches a realistic usage scenario for our technique.
\end{comment} 
\begin{figure}
\small
\[
\begin{array}{lll}
{\rm Cell \ type} \ \gamma & := & \texttt{num} \ |\  \texttt{string} \\
{\rm Primitive \ type} \ \beta & := & \gamma \ |\  \texttt{bool}\ |\ \texttt{cols} \\
{\rm Table \ type} \ \texttt{tbl} & :=&  \{l_1: \gamma_1, ... , l_n: \gamma_n \} \ \ (\texttt{\small row <: tbl})\\ 
{\rm Type} \ \tau & := &  \beta \ |  \ \texttt{tbl}  \ |\  \tau_1 \rightarrow \tau_2 \ | \  \tau_1 \times \tau_2  \\
\end{array}
\]
\vspace{-0.1in}
\caption{Types used in components; {\tt cols} represents a list of strings where each string is a column name in some table.}\label{fig:types}
\vspace{-0.2in}
\end{figure}

\section{Hypotheses as Refinement Trees}

\begin{figure}[t]
\[
\begin{array}{lll}
{\rm Term} \ t & := & {\rm const} | \ y_i \ | \ \comp(t_1, ..., t_n) \ (\comp \in \comps_v) \\
{\rm Qualifier} \ \qual & := &  (x, \tab) \ | \ \lambda y_1, \ldots y_n. \ t \\
{\rm Hypothesis} \ \hypo & := &   (?_i: \tau) \ | \ (?_i:\tau)@\qual \\
& & | \ ?_i^{\comp}(\hypo_1,...,\hypo_n) \ \ (\comp \in \comps_\tab)
\end{array}
\]
\vspace{-0.1in}
\caption{Context-free grammar for hypotheses}\label{fig:hypo}
\end{figure}

Before we can describe our synthesis algorithm, we first introduce  \emph{hypotheses} that represent partial programs with unknown expressions (i.e., holes). More formally,  hypotheses $\hypo$ are defined by the  grammar presented in Figure~\ref{fig:hypo}. 
In the simplest form, a hypothesis $(?_i: \tau)$   represents an unknown expression of type $\tau$. 
More complicated hypotheses are constructed using table transformation components $\comp \in \comps_\tab$. In particular, if $\comp = (f, \tau, \phi) \in \comps_\tab$, a hypothesis of the form $?_i^\comp(\hypo_1, \ldots, \hypo_n)$  represents an expression $f(e_1, \ldots, e_n)$.

During the course of our synthesis algorithm, we will progressively fill the holes in the hypothesis with concrete expressions. For this reason, we also allow hypotheses of the form $(?_i: \tau)@\qual$ where \emph{qualifier} $\qual$ specifies the term that is used to fill hole $?_i$. Specifically, if $?_i$ is of type {\tt tbl}, then its corresponding qualifier has the form $(x, \tab)$, which means that $?_i$ is instantiated with input variable $x$, which is in turn bound to table $\tab$ in the input-output example provided by the user. On the other hand, if $?_i$ is of type $(\tau_1 \times \ldots \times \tau_n) \rightarrow \tau$, then then the qualifier must be a first-order function $\lambda y_1, \ldots y_n. t$ constructed using  components~$\comps_v$.~\footnote{We view constants as a special case of first-order functions.}

Since our synthesis algorithm starts with the most general hypothesis and progressively makes it more specific, we now define what it means to \emph{refine} a hypothesis:

\begin{definition}({\bf Hypothesis refinement})
Given two hypotheses $\hypo, \hypo'$, we say that $\hypo'$ is a refinement of $\hypo$ if it can be obtained by replacing some subterm $?_i:\tau$ of $\hypo$ by $?_i^\comp(\hypo_1, \ldots, \hypo_n)$ where $\comp = (f, \tau' \rightarrow \tau, \phi) \in \comps_\tab$.
\end{definition}

\begin{figure}[t]
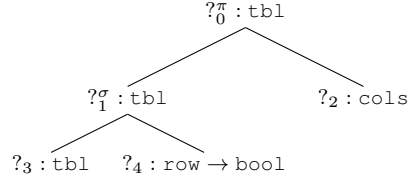

\vspace{-0.1in}
%\Tree [.$?_0:\tau_0$ ]
%\Tree [.$?_0^{\project}:\tau_0$ $?_1:\tau_1$  $?_2:\tau_2$ ]
\footnotesize
%\small
\Tree [.$?_0^{\project}:\texttt{tbl}$ [.$?_1^{\select}:\texttt{tbl}$ $?_3:\texttt{tbl}$ $?_4:{\texttt{row} \rightarrow \texttt{bool}}$ ] $?_2:\texttt{cols}$ ]
%\Tree [.$?_0^{\project}:\tau_0$ $?_1:\tau_1$@$(x_1,$\tab$)$  $?_2:\tau_2$ ]
\caption{Representing hypotheses as refinement trees}\label{fig:tree}
\vspace{-0.1in}
\end{figure}

In other words, a hypothesis $\hypo'$ refines another hypothesis $\hypo$ if it makes it more constrained. 

\begin{example}
The hypothesis $\hypo_1= ?_0^\sigma(?_1: \texttt{tbl}, ?_2: \texttt{row} \rightarrow \texttt{ bool})$ is a refinement of $\hypo_0 = ?_0:\texttt{tbl}$ because $\hypo_1$ is more specific than $\hypo_0$. 
In particular, $\hypo_0$ represents any arbitrary expression of type {\tt tbl}, whereas $\hypo_1$ represents expressions whose top-level construct is a selection.
\end{example}

Since our synthesis algorithm starts with the hypothesis $?_0: \texttt{tbl}$ and iteratively refines it, we will represent hypotheses using \emph{refinement trees}~\cite{myth}. Effectively, a refinement tree corresponds to the \emph{abstract syntax tree (AST)}  for the hypotheses from Figure~\ref{fig:hypo}. In particular, note that internal nodes labeled $?_i^\chi$ of a refinement tree represent hypotheses whose top-level construct is  $\chi$. If an internal node  $?_i^\chi$ has children labeled with unknowns $?_{j}, \ldots, ?_{j+n}$, this means that hypothesis $?_i$ was refined to $\chi(?_j, \ldots, ?_{j+n})$. Intuitively, a refinement tree captures the \emph{history} of refinements that occur as we search for the desired program.

\begin{figure}[t]
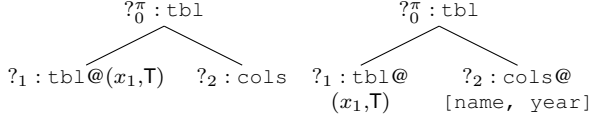

\footnotesize
\Tree [.$?_0^{\project}:\texttt{tbl}$ $?_1:\texttt{tbl}$@$(x_1,$\tab$)$  $?_2:\texttt{cols}$ ]
\Tree [.$?_0^{\project}:\texttt{tbl}$ $?_1:\texttt{tbl}$@\\$(x_1,$\tab$)$  $?_2:\texttt{cols}$@\\{\tt [name, year]} ]
\caption{A sketch (left) and a complete program (right)}\label{fig:complete}
\end{figure}

\begin{figure*}
\[
\begin{array}{llllll}
\begin{array}{llllllllll}
\peval{(?_i:\tau)} = ?_i  & \qquad \qquad \qquad 
\peval{(?_i:\tau)@(x, \tab)} = \tab & \qquad  \qquad \qquad 
\peval{(?_i:\tau)@t}  =  t 
\end{array}   
\smallskip 
\\ 
\peval{?_i^\chi(\hypo_1, \ldots, \hypo_n)} =  
\left \{ 
\begin{array}{ll} 
\comp(\peval{\hypo_1}, \ldots, \peval{\hypo_n}) & {\rm if}\  \exists i \in[1,n]. \ {\textsc{Partial}}(\peval{\hypo_i}) \\
\eval{\comp(\peval{\hypo_1}, \ldots, \peval{\hypo_n})} & {\rm otherwise}
\end{array}
\right .
\end{array}
\]
\vspace{-0.1in}
\caption{Partial evaluation of hypothesis. We write {\sc Partial}($\peval{\hypo}$) if $\peval{\hypo}$ contains at least one question mark.}\label{fig:partial-eval}
\vspace{-0.1in}
\end{figure*}

\begin{example}
Consider the refinement tree from Figure~\ref{fig:tree}, and suppose that $\pi, \sigma$ denote the standard
projection and selection operators in relational algebra. This refinement tree represents the partial program
$\pi(\sigma(?, ?), ?)$.% where $?$'s represent unknown expressions. 
The refinement tree also captures the search history in our 
synthesis algorithm. Specifically, it shows that our initial hypothesis was $?_0$, which then got refined to $\pi(?_1)$, which in turn got refined to $\pi(\sigma(?_3, ?_4), ?_2)$.
\end{example}

As mentioned in Section~\ref{sec:intro}, our approach decomposes the synthesis task into two separate \emph{sketch generation} and \emph{sketch completion} phases. We define a \emph{sketch} to be a special kind of hypothesis where there are no unknowns of type {\tt tbl}.

\begin{definition}{(\bf Sketch)}
A \emph{sketch} is a special form of hypothesis where all leaf nodes of type {\tt tbl} have a corresponding qualifier of the form $(x, \tab)$. 
\end{definition}

In other words, a sketch completely specifies the  table transformers used in the target program, but the first-order functions supplied as arguments to the table transformers are yet to be determined.

\begin{example}
Consider the refinement tree from Figure~\ref{fig:tree}. This hypothesis is not a sketch because there is a leaf node (namely $?_3$) of type {\tt tbl} that does not have a corresponding qualifier. On the other hand, the refinement tree shown in Figure~\ref{fig:complete} (left) is a sketch and corresponds to the partial program $\pi(x_1, ?)$ where $?$ is a list of column names. Furthermore, this sketch states that variable $x_1$ corresponds to table $\tab$ from the input-output example.
\end{example}

\begin{definition}{(\bf Complete program)}
A \emph{complete program}  is a hypothesis where all leaf nodes are of the form $(?_i: \tau)@\qual$.
\end{definition}

In other words, a complete program fully specifies the expression represented by each $?$ in the hypothesis. For instance, a hypothesis that represents a complete program is shown in Figure~\ref{fig:complete} (right) and represents the relational algebra term $\lambda x_1. \pi_{\emph{name, year}}(x_1)$. 

\begin{figure}[b]
\begin{tabular}{ll}
     \begin{tabular}{c c c c} 
     $\tab_1$ \\
     \hline
    \rowcolor{lightgoldenrodyellow}
     id & name & age & GPA \\ [0.5ex] 
     \hline\hline
    \rowcolor{babygreen}
     1 & Alice & 8 & 4.0 \\ 
     \hline
    \rowcolor{babypurple}
     2 & Bob & 18 & 3.2 \\
     \hline
    \rowcolor{babygreen}
     3 & Tom & 12 & 3.0 \\
     \hline
    \end{tabular}
&
     \begin{tabular}{c c c c} 
     $\tab_2$  \\
     \hline
    \rowcolor{lightgoldenrodyellow}
     id & name & age & GPA \\ [0.5ex] 
     \hline\hline
    \rowcolor{babygreen}
     2 & Bob & 18 & 3.2 \\
     \hline
    \rowcolor{babypurple}
     3 & Tom & 12 & 3.0 \\
     \hline
    \end{tabular}

\end{tabular}
\caption{Tables for Example~\ref{ex:partial}}\label{fig:partial-tables}
\end{figure}

As mentioned in Section~\ref{sec:intro}, our synthesis procedure relies on performing partial evaluation. Hence, we define a function $\peval{\hypo}$, shown in Figure~\ref{fig:partial-eval},  for partially evaluating hypothesis $\hypo$. Observe that, if $\hypo$ is a complete program, then $\peval{\hypo}$ evaluates to a concrete table. Otherwise, $\peval{\hypo}$ returns a partially evaluated hypothesis. We write {\sc Partial}($\peval{\hypo}$) if $\peval{\hypo}$ does not evaluate to a concrete term (i.e., contains question marks).

\begin{figure}[t]
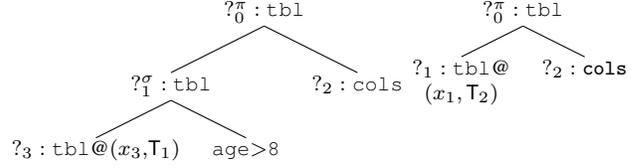

\vspace{-0.1in}
\footnotesize
\Tree [.$?_0^{\project}:\texttt{tbl}$ [.$?_1^{\select}:\texttt{tbl}$ $?_3:\texttt{tbl}$@$(x_3,$\tab_1$)$ \texttt{age$>$8} ] $?_2:\texttt{cols}$ ]
\Tree [.$?_0^{\project}:\texttt{tbl}$ $?_1:\texttt{tbl}$@\\$(x_1,\tab_2)$ $?_2:{\tt cols}$ ]
\caption{Partial evaluation on hypothesis from Figure~\ref{fig:tree}; {\tt \small age$>$8}
stands for $?_4:\texttt{\small row} \rightarrow \texttt{\small bool}@\lambda x. \ (x.\texttt{\small age}>8)$.}\label{fig:treePe}
\vspace{-0.2in}
\end{figure}

\begin{example}\label{ex:partial}
Consider  hypothesis $\hypo$  on the left-hand side of Figure~\ref{fig:treePe}, where $\tab_1$ is Table 1 from Figure~\ref{fig:partial-tables}. The refinement tree on the right-hand-side of Figure~\ref{fig:treePe} shows the result of partially evaluating $\hypo$, where $\tab_2$ is Table 2 from Figure~\ref{fig:partial-tables}.
\end{example}

\begin{comment}
 \begin{tabular}{p{4cm}p{4cm}}
      \hspace{-0.8cm}
     \begin{tabular}{||c c c c||} 
     \hline
     id & name & age & GPA \\ [0.5ex] 
     \hline\hline
     1 & Alice & 8 & 4.0 \\ 
     \hline
     2 & Bob & 18 & 3.2 \\
     \hline
     3 & Tom & 12 & 3.0 \\
     \hline
    \end{tabular}
        & \hspace{-0.8cm}
     \begin{tabular}{||c c c c||} 
     \hline
     id & name & age & GPA \\ [0.5ex] 
     \hline\hline
     2 & Bob & 18 & 3.2 \\
     \hline
     3 & Tom & 12 & 3.0 \\
     \hline
    \end{tabular}
\end{tabular}
\end{comment}

\begin{comment}
%\begin{figure}
\begin{table}
\parbox{.45\linewidth}{
\centering
     \begin{tabular}{||c c c c||} 
     \hline
     id & name & age & GPA \\ [0.5ex] 
     \hline\hline
     1 & Alice & 8 & 4.0 \\ 
     \hline
     2 & Bob & 18 & 3.2 \\
     \hline
     3 & Tom & 12 & 3.0 \\
     \hline
    \end{tabular}
\caption{$\tab_1$}
}
\hfill
\parbox{.45\linewidth}{
\centering
     \begin{tabular}{||c c c c||} 
     \hline
     id & name & age & GPA \\ [0.5ex] 
     \hline\hline
     2 & Bob & 18 & 3.2 \\
     \hline
     3 & Tom & 12 & 3.0 \\
     \hline
    \end{tabular}
\caption{$\tab_2$}
}
\caption{$\tab_2$}
\end{table}
%\caption{Tables for Example~\ref{ex:partial}}\label{fig:partial-tables}
%\end{figure}
\end{comment}

\section{Synthesis Algorithm}\label{sec:alg}

In this section, we describe the high-level structure of our synthesis algorithm, leaving the discussion of SMT-based deduction and sketch completion to the next two sections.

Figure~\ref{fig:synthesis} illustrates the main ideas underlying our  synthesis algorithm.  The idea is to maintain a priority queue of hypotheses, 
which are either converted into a sketch or refined to a more specific hypothesis during each iteration. Specifically, the synthesis procedure picks the most promising hypothesis $\hypo$ according to some heuristic cost metric (explained in Section~\ref{sec:impl}) and asks the deduction engine if $\hypo$ can be successfully converted into a sketch. If the deduction engine refutes this conjecture, we then discard $\hypo$ but add all possible (one-level) refinements of $\hypo$  into the worklist. Otherwise, we convert hypothesis $\hypo$ into a sketch $\sketch$ and try to complete it using the \emph{sketch completion engine}. 

\tikzstyle{block} = [rectangle, draw, text width=4.8em, text centered, rounded corners, minimum height=3em,line width=0.3mm]
\begin{figure}[!b]
\vspace{-0.1in}
\centering
{\small
\begin{tikzpicture}[node distance = 2cm, auto,thick,scale=0.9, every node/.style={scale=0.9}]
  % Draw nodes
  \node [block] (refinement) {Hypothesis Refinement};
  \node [block,right of=refinement, node distance=3.6cm] (deduction) {SMT-based Deduction};
  \node [block,below of=deduction] (completion) {Sketch Completion};
  \node [block,right of=completion, node distance=3cm] (program) {Program};

  \draw [line width=0.3mm,->] (refinement) -- (deduction);
  \draw [line width=0.3mm,->] (completion) -- (program);
  \draw [line width=0.3mm,->] (deduction) -- (completion);
  \draw [line width=0.3mm,->] (completion) to [out=-180,in=270,looseness=1] (refinement);
  \draw [line width=0.3mm,->] (deduction) to [out=90,in=90,looseness=0.6] (refinement);

  \node at (3.9,-1) {\textcolor{camouflagegreen}{\LARGE \cmark}};
  \node at (5.1,-1.7) {\textcolor{camouflagegreen}{\LARGE \cmark}};
  \node at (1.8,0.85) {\textcolor{red}{\LARGE \xmark}};
  \node at (3.1,-1) {sketch};
  \node at (1.8,0.25) {candidate};
  \node at (1.8,-0.2) {sketch};
  \node at (1.5,-1.55) {\textcolor{red}{\LARGE \xmark}};
\end{tikzpicture}
}
\caption{Illustration of the top-level synthesis algorithm}\label{fig:synthesis}
\vspace{-0.1in}
\end{figure}
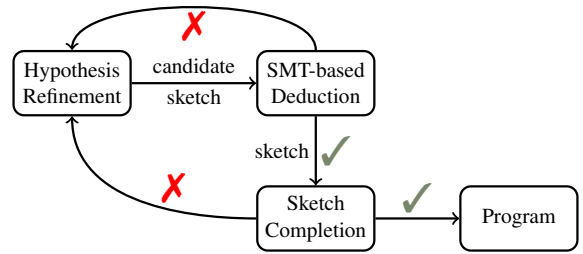

% \begin{figure}[!b]
%     \centering
%     \includegraphics[scale=0.4]{morpheus2.pdf}
%     \caption{Illustration of the top-level synthesis algorithm}\label{fig:synthesis}
% \end{figure}

Our top-level synthesis algorithm is presented in 
Algorithm~\ref{alg:synthesis}. Given an example $\ex$ and
a set of components $\comps$, {\sc Synthesize} either returns a
complete program that satisfies $\ex$ or yields $\bot$, meaning that
no such program exists.

In more detail, the {\sc Synthesize} procedure maintains a priority queue $\worklist$ of all hypotheses. Initially, the only hypothesis in $\worklist$ is $?_0$, which represents any possible program. In each iteration of the while loop (lines 5--18), we pick a hypothesis $\hypo$ from $\worklist$ and invoke the {\sc Deduce} procedure (explained later) to check if $\hypo$ can be directly converted into a sketch by filling  holes of type {\tt tbl} with the input variables. Note that our deduction procedure is sound but, in general, not complete: In particular, since the specifications of components can be imprecise, the deduction procedure can return $\top$ (i.e., true) even though no valid completion of the sketch exists. However, {\sc Deduce} returns $\bot$ only when the current hypothesis requires further refinement.

\begin{algorithm}[t]
\caption{Synthesis Algorithm}\label{alg:synthesis}
{%\small 
\begin{algorithmic}[1]
%\vspace{0.05in}
\Procedure{Synthesize}{$\sample, \comps$}
%\vspace{0.05in}
\State {\rm \bf input:} Input-output example $\ex$  and components $\comps$

%\State \ \ \ \ \ \ \ \ \ \ \ \ components $\comps$, and tests $\tests$
\State {\rm \bf output:} Synthesized program or $\bot$ if failure
\vspace{0.05in}

\State  $\worklist :=$  \{$?_0$:{\tt tbl}\} \Comment{ \ Init worklist}

\vspace{0.05in}
\While {$\worklist \neq \emptyset$} 
    \State {\rm choose} $\hypo \in \worklist$; 
    \State $\worklist := \worklist \backslash \{\hypo\}$
    \If{{\sc Deduce}($\hypo,\sample) = \bot$} \Comment{\ Contradiction}
    \State {\rm \bf goto} refine;
    \EndIf
    \State \Comment{\ No contradiction}
        \For{$\sketch \in$ {\sc Sketches}($\hypo, \sample_{in})$}
            \State $\prog$ := {\sc Fillsketch}($\sketch, \sample$) 
            %\State \Comment{Sketch completion}
            \For{$p \in \prog$}
                \If{{\sc check}($p,\sample$)}
                    \Return $p$
                \EndIf
            \EndFor
        \EndFor
     \vspace{0.1in}
    \State {\bf refine:} \Comment{Hypothesis refinement}
    \vspace{0.05in}
    \For{$\comp \in \comps_\tab$, $(?_i$: {\tt tbl}) $\in$ {\sc Leaves}($\hypo$)  }
    \State $\hypo' := \hypo[?_{j}^{\comp}(?_j:\vec{\tau}) / ?_i]$
    \State $\worklist := \worklist \cup \hypo' $
\EndFor
\EndWhile
\vspace{0.05in}
\State \Return $\bot$
\EndProcedure
\end{algorithmic}
}
\end{algorithm}

If {\sc Deduce} does not find a conflict, we then convert the current hypothesis $\hypo$ into a set of possible sketches (line 11). The function {\sc Sketches} used at line 11 is presented using  inference rules in Figure~\ref{fig:sketch-rules}. Effectively, we convert hypothesis $\hypo$ into a sketch by replacing each hole of type {\tt tbl} with one of the input variables $x_j$, which corresponds to table $\tab_j$ in the input-output example.

After we obtain a candidate sketch, we try to complete it using the
call to {\sc FillSketch} at line 12 (explained in Section~\ref{sec:completion}). {\sc FillSketch} returns a \emph{set} of complete programs $\prog$ such that each $p \in \prog$ is valid with respect to our deduction procedure. However, as our deduction  procedure is incomplete, $p$ may not satisfy the input-output examples. Hence, we only return $p$ as a solution if $p$ satisfies $\ex$  (line 14).

Lines 16-18 of Algorithm~\ref{alg:synthesis} perform \emph{hypothesis refinement}. The idea behind hypothesis refinement is to replace one of the holes of type {\tt tbl} in $\hypo$ with a component from $\comps_\tab$, thereby obtaining a more specific hypothesis. Each of the refined hypotheses is added to the worklist and possibly converted into a sketch in future iterations.

\begin{figure}[!t]
\vspace{-0.2in}
{%\small 
\begin{center}
\[
\begin{array}{cr}
\irule{
\begin{array}{c}
 \tab_j \in \tab_{in} \\
 \hypo = (?_i:\texttt{tbl}) 
 \end{array}
}
{
  \hypo @(x_j, \tab_j) \in \emph{Sketches}(\hypo, \vec{\tab}_{in})  
} &  ({\rm 1}) \smallskip \ \\

\irule{
\begin{array}{c}
 \tau_i \neq \texttt{tbl} \\
 \hypo = ?_i:\tau_i
 \end{array}
}
{
  \hypo \in \emph{Sketches}(\hypo, \vec{\tab}_{in})
} &  ({\rm 2}) \smallskip \ \smallskip \\

\irule {
\begin{array}{c}
 \hypo = ?_i^\comp(\hypo_1,...,\hypo_n)\\
 \hypo_i' \in \emph{Sketches}(\hypo_i, \vec{\tab}_{in})
 \end{array}
}
{ ?_i^\comp(\hypo_1',...,\hypo_n') \in \emph{Sketches}(\hypo, \vec{\tab}_{in})} & ({\rm 3})

\end{array}
\]
\end{center}
}
\vspace{-0.1in}
\caption{Converting a hypothesis into a sketch.}
\vspace{-0.1in}
\label{fig:sketch-rules}
\end{figure}

\begin{comment}
\subsection{Predicate generation}
Problem definition:  Given $\select_{p}(T) : \constraint$ where T is the input table and $\constraint$ is the output constraint, compute predicate $p$ that satisfies the transformation.
\begin{itemize}
\item Obtain all possible predicates: $p_1, ..., p_n$
\item Obtain a list of table $T_1, ..., T_n$ by applying the above predicates
to the input table. e.g., for component $\select$, $\select_{p_i}(T_{in}) = T_i$
\item Figure out a list of predicates that are likely to be involved in the constraint 
by solving the following MAX SAT formula where $\constraint$ and $SuperTable(T_i, T)$ are
hard and soft constraints, respectively:
\[
SuperTable(T_1, T) \land ... \land SuperTable(T_n,T) \land \constraint
\]
\item Generate a list of actual predicates by directly picking from predicates computed by previous step or a formula comprised by a subset of them: $p_1, ..., p_j$, $p_i$ is the right predicate if it satisfies the following constraint:
\[
SAT(T_i = T) \land \constraint
\]

\end{itemize}
\end{comment}

%\section{Formalization of Algorithm}\label{sec:alg}

\section{SMT-based Deduction}\label{sec:deduction}

We now turn to the {\sc Deduce} procedure used in Algorithm~\ref{alg:synthesis}. 
The key idea here is to generate an SMT formula that
corresponds to the specification of the current sketch and to check whether the input-output example satisfies this specification.

\begin{table*}
\vspace{-0.1in}
\centering
\newcolumntype{A}{ >{\arraybackslash} m{0.41\textwidth} }
\newcolumntype{B}{ >{\centering\arraybackslash} m{0.085\textwidth} }
\newcolumntype{E}{ >{\centering\arraybackslash} m{0.24\textwidth} }
\newcolumntype{D}{ >{\centering\arraybackslash} m{0.03\textwidth} }
\scalebox{0.9}{
\begin{tabular}{| D | B | A | E |}
   \hline
    Lib & Component & \centering Description & Specification \\
\hline
    \multirow{3}{*}{\begin{turn} {90}\makecell{tidyr}\end{turn}}
&  spread & Spread a key-value pair across multiple columns. &  $\tab_{out}.\rows \leq \tab_{in}.\rows$ \newline $\tab_{out}.\columns \geq \tab_{in}.\columns$ \\
\cline{2-4}
&  gather & Takes multiple columns and collapses into key-value pairs, duplicating all other columns as needed.& $\tab_{out}.\rows \geq \tab_{in}.\rows$ \newline $\tab_{out}.\columns \leq \tab_{in}.\columns$ \\
\cline{2-4}
%&  separate & Turns a single  column into multiple columns. & $\tab_{out}.\rows = \tab_{in}.\rows$ \newline  $\tab_{out}.\columns = \tab_{in}.\columns + 1$\\
%\cline{2-4}
    \hline
    \multirow{3}{*}{\begin{turn} {90}\makecell{dplyr}\end{turn}}
&  select & Project a subset of columns in a data frame. & $\tab_{out}.\rows = \tab_{in}.\rows$ \newline $\tab_{out}.\columns < \tab_{in}.\columns$ \\
\cline{2-4}
&  filter & Select a subset of rows in a data frame. &$\tab_{out}.\rows < \tab_{in}.\rows$\newline$\tab_{out}.\columns = \tab_{in}.\columns$\\ 
\cline{2-4}
%&  mutate & Add new columns to the data frame based on operations on existing columns. & $\tab_{out}.\rows = \tab_{in}.\rows$\newline$\tab_{out}.\columns = \tab_{in}.\columns + 1$ \\
%\cline{2-4}
\hline
%    \multirow{2}{*}{\begin{turn} {90}\makecell{built-in}\end{turn}}
%&  cbind & Take a sequence of vector, matrix or data frame arguments and combine by columns & $\tab_{out}.\rows = \tab_{in}^{1}.\rows$\newline$\tab_{out}.\rows = \tab_{in}^{2}.\rows$\newline$\tab_{out}.\columns = \tab_{in}^{1}.\columns + \tab_{in}^{2}.\columns$ \\
%\cline{2-4}
%&  rbind & Take a sequence of vector, matrix or data frame arguments and combine by rows& $\tab_{out}.\columns = \tab_{in}^{1}.\columns$ \newline $\tab_{out}.\columns = \tab_{in}^{2}.\columns$ \newline $\tab_{out}.\rows = \tab_{in}^{1}.\rows + \tab_{in}^{2}.\rows$   \\
%\cline{2-4}
%\hline
\end{tabular}
}
\caption{Sample specifications of a few components}\label{table:spec}
\vspace{-0.1in}
\end{table*}

\paragraph{Component specifications.} We use the specifications of
individual components to derive the overall specification for a given
hypothesis. As mentioned earlier, these specifications need not be
precise and can, in general, overapproximate the behavior of the
components. For instance, Table~\ref{table:spec} shows sample
specifications for a subset of methods from two popular R
libraries. Note that these sample specifications do not fully
capture the behavior of each component and only describe the
relationship between the number of rows and columns in the input and
output tables.~\footnote{The actual specifications used in our
  implementation are slightly more involved. In
  Section~\ref{sec:eval}, we compare the performance of \toolname
  using two different specifications.}
For example, consider the {\tt filter} function from the {\tt dplyr}
library for selecting a subset of the rows that satisfy a given
predicate in the data frame. The specification of {\tt filter}, which
is effectively the selection operator $\sigma$ from
relational algebra, is given by:
\[
\tab_{out}.\rows < \tab_{in}.\rows \land \tab_{out}.\columns = \tab_{in}.\columns
\]
In other words, this specification expresses that the table obtained
after applying the {\tt filter} function contains fewer rows but the
same number of columns as the input table.~\footnote{In principle, the
  number of rows may be unchanged if the predicate does not match any row. However, we need not consider this case since there is a simpler program without {\tt filter}  that satisfies the  example.}

\begin{figure}
{\small 
\[
\begin{array}{cl}
\toConstraint(\hypo_i) & =  \alpha(\peval{\hypo_i})[?_i/x] \ {\rm if} \ \neg {{\small \textsc{Partial}}}(\peval{\hypo_i})  \\
\toConstraint(\hypo_i) & =  \top  \qquad \quad  {\rm else \  if} \ {\textsc{Isleaf}}(\hypo_i) \\
\toConstraint(?_0^{\comp}(\hypo_1,...,\hypo_n)) & =  \bigwedge\limits_{1\leq i \leq n}\toConstraint(\hypo_i)\land \constraint_\chi[?_0 /y, \vec{?_i}/\vec{x_i}]
\end{array}
\]
}
\caption{Constraint generation for hypotheses.  $?_i$ denotes the root variable of $\hypo_i$ and  the specification of  $\comp$ is $\phi_{\comp}$.}\label{fig:cst-gen}
\end{figure}

\paragraph{Generating specification for hypothesis.} Given a hypothesis $\hypo$, we need to generate the specification for $\hypo$ using the specifications of the individual components used in $\hypo$. Towards this goal, the function $\toConstraint(\hypo)$ defined in Figure~\ref{fig:cst-gen} returns the specification of hypothesis $\hypo$. 

In the simplest case,  $\hypo_i$ corresponds to a complete program (line 1 of Figure~\ref{fig:cst-gen})~\footnote{Recall that the {\sc Deduce} procedure will also be used during sketch completion. While $\hypo$ can never be a complete program when called from line 8 of the {\sc Synthesize} procedure (Algorithm~\ref{alg:synthesis}), it can be a complete program when {\sc Deduce} is invoked through the sketch completion engine.}. In this case, we evaluate the hypothesis to a  table $\tab$ and obtain $\Phi(\hypo_i)$ as the ``abstraction" of $\tab$. In particular, the \emph{abstraction function} $\alpha$ used in Figure~\ref{fig:cst-gen} takes as input a concrete table $\tab$ and returns a constraint describing that table. In general, the definition of the abstraction function $\alpha$ depends on the granularity of the component specifications. For instance, if our component specifications only refer to the number of rows and columns, then a suitable abstraction function for an $m \times n$ table would yield $x.\rows = m \land x.\columns = n$. In general, we assume variable $x$ is used to describe the input table of $\alpha$.

Let us now consider the second case in Figure~\ref{fig:cst-gen} where $\hypo_i$ is a leaf, but not a complete program. In this case, since we do not have any information about  what $\hypo_i$ represents, we return $\top$ (i.e., \emph{true}) as the specification.

Finally, let us consider the case where the hypothesis is of the form $?_0^\comp(\hypo_1, \ldots, \hypo_n)$. In this case, we first recursively infer the specifications of sub-hypotheses $\hypo_1, \ldots, \hypo_n$. Now suppose that the specification of $\comp$ is given by $\phi_\comp(\vec{x}, y)$, where $\vec{x}$ and $y$ denote $\comp$'s inputs and output respectively. If the root variable of each hypothesis $\hypo_i$ is given by $?_i$, then the specification for the overall hypothesis is obtained as: 
\[
\bigwedge\limits_{1\leq i \leq n}\toConstraint(\hypo_i)\land \constraint_\chi[?_0 /y, \vec{?_i}/\vec{x_i}]
\]

\begin{example}
Consider  hypothesis $\hypo$ from Figure~\ref{fig:tree}, and suppose that the specifications for relational algebra operators $\pi$ and $\sigma$ are the same as {\tt select} and {\tt filter} from Table~\ref{table:spec} respectively. Then, $\Phi(\hypo)$ corresponds to the following Presburger arithmetic formula:
\[
\begin{array}{c}
?_1.\rows < ?_3.\rows \ \land \  ?_1.\columns = ?_3.\columns \ \land \\
?_0.\rows = ?_1.\rows \ \land \ ?_0.\columns < ?_1.\columns
\end{array}
\]
Here, $?_3, ?_0$ denote the input and output tables respectively, and $?_1$ is the intermediate table obtained after selection.
\end{example}

\paragraph{Deduction using SMT.} Algorithm~\ref{alg:deduce} presents our deduction algorithm using the constraint generation function $\Phi$ defined in Figure~\ref{fig:cst-gen}. Given a hypothesis $\hypo$ and input-output example $\ex$, {\sc Deduce} returns $\bot$ if $\hypo$ does not correspond to a valid sketch. In other words, {\sc Deduce}$(\hypo, \ex) = \bot$ means that we cannot obtain a program that satisfies the input-output examples by replacing  holes with inputs.

As shown in Algorithm~\ref{alg:deduce}, the {\sc Deduce} procedure generates a constraint $\psi$ and checks its satisfiability using an SMT solver. If $\psi$ is unsatisfiable,  hypothesis $\hypo$ cannot be unified with the input-output example and can therefore be rejected.

Let us now consider the construction of SMT formula $\psi$ in Algorithm~\ref{alg:deduce}. First, given a hypothesis $\hypo$, the corresponding sketch must map each of the unknowns of type {\tt tbl} to one of the arguments. Hence, the constraint $\varphi_\emph{in}$ generated at line 5 indicates that each leaf with label $?_j$ corresponds to some argument $x_i$. Similarly, $\varphi_\emph{out}$ expresses that the root variable of hypothesis $\hypo$ must correspond to the return value $y$ of the synthesized program. Hence, the constraint $\toConstraint(\hypo) \land \varphi_\emph{in} \land \varphi_\emph{out}$ expresses the specification of the sketch in terms of variables $x_1, \ldots, x_n, y$. 

Now,  to check if $\hypo$ is unifiable with  example $\ex$, we must also generate constraints that describe each table $\tab_{in}^i$ in terms of $x_i$ and $\tab_{out}$ in terms of $y$. Recall from earlier that the abstraction function $\alpha(\tab)$ generates an SMT formula describing $\tab$ in terms of variable $x$. Hence,  the constraint

\[
\bigwedge\limits_{\tab_i \in \sample_{in}}(\abs(\tab_i)[x_i/x]) \land \abs(\tab_{out})[y/x]
\]
expresses that each $\tab_{in}^i$ must correspond to $x_i$ and $\tab_{out}$ must correspond to variable $y$. Thus, the unsatisfiability of formula $\psi$ at line 7 indicates that hypothesis $\hypo$ can be rejected.

\begin{algorithm}[t]
\caption{SMT-based Deduction Algorithm}\label{alg:deduce}
\begin{algorithmic}[1]
\vspace{0.05in}
\Procedure{Deduce}{$\hypo, \sample$}
\vspace{0.05in}
\State {\rm \bf input:} Hypothesis $\hypo$, input-output example $\ex$
\State {\rm \bf output:}  $\bot$ if  cannot be unified with $\ex$; $\top$ otherwise
\vspace{0.05in}
\State $\sketch := \{?_j \ | \ ?_j:\texttt{tbl} \in \textsc{leaves}(\hypo)\}$
\State $\varphi_{in} := \bigwedge\limits_{?_j \in \sketch }\bigvee\limits_{1\leq i \leq |\sample_{in}|}(?_j = x_i)$
\vspace{0.05in}
\State $\varphi_{out}$ $:=$ ($y=${\sc Rootvar}($\hypo$))
%\State $\constraint'$ := $\toConstraint(\hypo) \land \varphi_{in} \land \varphi_{out} \land \bigwedge\limits_{\tab_i \in \sample_{in}}\abs(\tab_i)[x_i/x] \land \abs(\tab_{out})[y/x]$
\vspace{0.05in}
\State $\psi$ $:= $
$\left (  
\begin{array}{c}
\toConstraint(\hypo) \land \varphi_{in} \land \varphi_{out} \land \\
\bigwedge\limits_{\tab_i \in \sample_{in}}(\abs(\tab_i)[x_i/x]) \land \abs(\tab_{out})[y/x]
\end{array}
\right ) 
$
\vspace{0.05in}
\State \Return {\sc sat}($\psi$)

\EndProcedure

\end{algorithmic}
\end{algorithm}

\begin{example}
Consider the hypothesis from Figure~\ref{fig:tree}, and suppose that the input and output tables are $\tab_1$ and $\tab_2$ from Figure~\ref{fig:partial-tables} respectively. The {\sc Deduce} procedure from Algorithm~\ref{alg:deduce} generates the following constraint $\psi$:
\[
\begin{array}{c}
?_1.\rows < ?_3.\rows \ \land \  ?_1.\columns = ?_3.\columns \ \land ?_0.\rows = ?_1.\rows \\  \ \land \ ?_0.\columns < ?_1.\columns \land 
x_1 = ?_3 \land y = ?_0 \ \land \\ x_1.\rows = 3 \land 
x_1.\columns = 4  \land   y.\rows = 2 \land y.\columns = 4
\end{array}
\]
Observe that $\toConstraint(\hypo) \land \varphi_{in} \land \varphi_{out}$ implies $y.\columns < x_1.\columns$, indicating that the output table should have fewer columns than the input table. Since we have $x_1.\columns = y.\columns$, constraint $\psi$ is unsatisfiable, allowing us to reject the hypothesis.
\end{example}

\section{Sketch Completion}\label{sec:completion}

Recall that the goal of sketch completion is to fill the remaining holes in the hypothesis with first-order functions constructed using components in $\comps_v$. For instance, consider the sketch $\pi(\sigma(x, ?_1), ?_2)$ where $\pi, \sigma$ are the familiar projection and selection operators from relational algebra. Now, in order to fill  hole $?_1$, we need to know the columns in table $x$. Similarly, in order to fill hole $?_2$, we need to know the columns in the intermediate table obtained using selection.

As this example illustrates, the vocabulary of first-order functions that can be supplied as arguments to table transformers often depends on the shapes (i.e., schemas) of the other arguments of type {\tt tbl}. For this reason, our sketch completion algorithm synthesizes the program \emph{bottom-up}, evaluating terms of type {\tt tbl} before synthesizing the other arguments. The concrete tables that are obtained by evaluating sub-terms of the sketch therefore determine the universe of constants that can be used in the  synthesis task.

At a high level, our sketch completion procedure synthesizes an argument of type $\tau$ by enumerating all inhabitants of type $\tau$. However, as argued earlier, the valid inhabitants of type $\tau$ are determined by a particular table. Hence, our sketch completion procedure performs \emph{``table-driven type inhabitation"}, meaning that it computes the inhabitants of a given type with respect to a concrete table.

\paragraph{Table-driven type inhabitation.} Before we can explain the full sketch completion procedure, we first discuss the notion of \emph{table-driven type inhabitation}: That is, given a type $\tau$ and a concrete table $\tab$, what are all valid inhabitants of $\tau$ with respect to the universe of constants used in $\tab$?

We formalize this variant of the type inhabitation problem using the inference rules shown in Figure~\ref{fig:inhabit}.  Specifically, these rules  derive judgments of the form $\Gamma  \vdash t \in \Omega(\tau, \tab)$ where $\Gamma$ is a type environment mapping variables to types. The meaning of this judgment is that, under type environment $\Gamma$, term $t$ is a valid inhabitant of type $\tau$ with respect to table $\tab$. Observe that we need the type environment $\Gamma$ due to the presence of function types: That is, given a function type $\tau_1 \rightarrow \tau_2$, we need  $\Gamma$ to enumerate  valid inhabitants of $\tau_2$.

\begin{figure}[!t]
\vspace{-0.1in}
{\small 
\begin{center}
\[
\begin{array}{cr}
\irule{
\begin{array}{c}
 \emph{type}(\tab) = \{l_1:\tau_1,...,l_n:\tau_n\} \\
% C_i \in \mathcal{P}([1,n]) \\
 c = [l_i \ | \ i \in C_i ] \ {\rm for} \  C_i \in \mathcal{P}([1,n]) 
% c = \{[l_i \ | \ i \in C_i ] \ | \ C_i \in \mathcal{P}([1,n])
 \end{array}
}
{
 \Gamma \vdash c \in \inhabit (\texttt{cols}, \tab)
} &  ({\rm Cols}) \\ \ \\

\irule{
\begin{array}{c}
 c \in \tab, \ \ \emph{type}(c) = \tau \\
 \tau \in \{\texttt{num, string}\} 
 \end{array}
}
{
  \Gamma \vdash c \in \inhabit (\tau, \tab)
} &  ({\rm Const}) \\ \ \\

\irule{
\begin{array}{c}
\Gamma \vdash x: \tau
 \end{array}
} 
{
\Gamma \vdash  x \in \inhabit (\tau, \tab)
} &  ({\rm Var}) \\ \ \\

\irule {
\begin{array}{c}
 \Gamma \vdash t_1 \in \inhabit(\tau_1, \tab)\\
 \Gamma \vdash t_2 \in \inhabit(\tau_2, \tab)
 \end{array}
}
{ \Gamma \vdash (t_1,t_2) \in \inhabit(\tau_1 \times \tau_2, \tab) } & ({\rm Tuple}) \\ \ \\

\irule {
\begin{array}{c}
 (f, \tau' \rightarrow \tau, \phi) \in \Lambda_v \\
 \Gamma \vdash t \in \inhabit (\tau', \tab)
 \end{array}
}
{ \Gamma \vdash f(t) \in \inhabit(\tau, \tab) } & ({\rm App}) \\ \ \\
\irule {
\begin{array}{c}
\tau = (\tau_1 \times \ldots \times \tau_n \rightarrow \tau') \\
\Gamma' = \Gamma \cup \{ x_1: \tau_1, \ldots x_n:\tau_n\} \\ %\ \ (\vec{x} \ \emph{fresh}) \\
\Gamma' \vdash  t \in \Omega(\tau', \tab)
 \end{array}
}
{ \Gamma \vdash (\lambda x_1, \ldots,  x_n. \ t) \in \inhabit(\tau, \tab) } & ({\rm Lambda})
\end{array}
\]
\end{center}
}
\caption{Table-driven type inhabitation rules.}
\label{fig:inhabit}
\vspace{-0.1in}
\end{figure}

Let us now consider the type inhabitation rules from Figure~\ref{fig:inhabit}, starting with the {\rm Cols} rule. Recall that the {\tt cols} type represents a list of strings, where each string is the name of a column in some table. Clearly, the universe of strings that can be used in any inhabitant of {\tt cols} depends on table $\tab$. Hence, the {\rm Cols} rule essentially generates all possible combinations of the column names used in $\tab$. 

Next, consider the ${\rm Const}$ rule from Figure~\ref{fig:inhabit} for synthesizing constants of type {\tt num} and {\tt string}.~\footnote{Recall from Section~\ref{sec:problem} that these are the only types of values that can appear in tables.} Given table $\tab$, we consider a constant $c$ to be an inhabitant of $\tau$ if it appears in table $\tab$. In the general case, this strategy of  considering only those constants that appear in table $\tab$ amounts to a heuristic for finitizing the universe of constants. However, this heuristic works quite well in practice and does not lead to a loss of completeness in many cases. For instance, consider the selection operator $\sigma$ from relational algebra, and suppose that the desired predicate is  $\texttt{age} > c$, where {\tt age} is a column and $c$ is a constant. Since our goal is to synthesize a program that satisfies the input-output example, we can always find another predicate $\texttt{age} > c'$ where $c'$ occurs in the table and the two programs are equivalent modulo the inputs.

%The intuition is that, if there exists a program $P$ using a constant $c$ that satisfies the input-output example, we can always find another program  $P'$ that uses another constant $c'$ such that $c'$ appears in table $\tab$. In other words, programs $P, P'$ are \emph{equivalent modulo inputs}, and, since our goal is to synthesize a program that satisfies the input-output examples, it suffices to only consider programs that refer to constants used in the input tables. \todo{State some sort of completeness theorem and make this precise in the proof.}

The  {\rm Var} rule is very simple and says that variable $x$ is an inhabitant of  $\tau$ if it has type $\tau$ according to $\Gamma$. 
The {\rm Tuple} rule is also straightforward, and says that $(t_1, t_2)$ is an inhabitant of $\tau_1 \times \tau_2$ if $t_1, t_2$ are inhabitants of $\tau_1$ and $\tau_2$ respectively.

The next rule {\rm App} is more interesting and allows us to generate richer terms  using components in $\Lambda_v$. In particular, if $f: \tau' \rightarrow \tau$ is a component in $\comps_v$ and $t$ is an inhabitant of $\tau'$, the {\rm App} rule says that $f(t)$ is  an inhabitant of  $\tau$. For instance, given an operator $\geq: \texttt{num} \times \texttt{num} \rightarrow \texttt{bool} \in \comps_v$, the {\rm App} rule allows us to construct a term such as $x \geq 10$.

Finally, consider the {\rm Lambda} rule for synthesize inhabitants of function types. Observe that this rule is  necessary because table transformers can be higher-order functions. Given a function type $(\tau_1 \times \ldots \times \tau_n) \rightarrow \tau'$, we first generate  fresh variables $x_1, \ldots, x_n$ of type $\tau_1, \ldots, \tau_n$  and add them to  $\Gamma$. We then synthesize the body of the function using the new type environment $\Gamma'$.

\begin{example}
Consider  table $\tab_1$ from Figure~\ref{fig:partial-tables} and the type environment $\Gamma: \{ x \mapsto \texttt{string} \}$. Assuming $\texttt{eq}: \texttt{string} \times \texttt{string} \rightarrow \texttt{bool}$ is a component in $\comps_v$,  we have $\texttt{eq}(x, \texttt{"Alice"}) \in \Omega(\texttt{bool}, \tab_1) $ using the {\rm{App, Const, Var}} rules. Similarly, $\lambda x. \texttt{eq}(x, \texttt{"Bob"})$ is also a valid inhabitant of $\texttt{string} \rightarrow \texttt{bool}$ with respect to $\tab_1$.
\end{example}

\begin{comment}
\begin{figure}[t]
\[
\begin{array}{lll}
{\rm Type} \ \tau & := & \emph{Num} \ | \ \emph{String} \ | \ \emph{Bool} \ | \ \emph{ColList} \ | \  \tau \times \tau  
\end{array}
\]
\caption{Types of operands used in value transformers $\comps_v$}\label{fig:type}
\end{figure}
\end{comment}

\begin{figure}[!t]
{\small 
\begin{center}
\[
\begin{array}{cr}
\irule{
\begin{array}{c}
\sketch = (?_i:\tau_i) \\
t \in \Omega(\tau_i, \tab, \emptyset) \\
{\textsc{Deduce}}(\sketch_f[\sketch@t/\sketch], \ex) \neq \bot
 \end{array}
}
{
 \sketch@t \in \complete_v(\sketch, \sketch_f, \ex, \tab)
} &  (1) \\ \ \\

\irule{
\begin{array}{c}
\sketch = (?_i, \texttt{tbl})@(x, \tab) \\
 \end{array}
}
{
 (\sketch, \tab) \in \complete_\tab(\sketch, \sketch_f, \ex)
} &  (2) \\ \ \\

\irule{
\begin{array}{c}
\sketch = ?_i^\comp(\vec{\hypo}: \texttt{tbl}, \vec{\hypo'}: \tau) \ \ \ (\tau \neq  \texttt{tbl})  \\
(\prog_j, \tab_j) \in \complete_\tab(\hypo_j, \sketch_f, \ex) \\
%\tab = \tab_1 \times \ldots \times \tab_n \\
\prog_j' \in \complete_v(\hypo_j', \sketch_f[\vec{\prog}/\vec{\hypo}], \ex, \tab_1 \times \ldots \times \tab_n) \\
{\textsc{Deduce}}(\sketch_f[\vec{\prog}/\vec{\hypo}, \vec{\prog'}/\vec{\hypo'}], \ex) \neq \bot \\
\prog^* = \sketch[\vec{\prog}/\vec{\hypo}, \vec{\prog'}/\vec{\hypo'}] \\
 \end{array}
}
{
 (\prog^*, \peval{\prog^*}) \in \complete_\tab(\sketch, \sketch_f, \ex)
} &  (3)  \\ \ \\

\irule{
\begin{array}{c}
 (\prog, \tab) \in \complete_\tab(\sketch, \sketch, \ex) 

 \end{array}
}
{
 \prog \in {\textsc{FillSketch}}(\sketch, \ex)
} &  (4) 

\end{array}
\]
\end{center}
}
\vspace{-0.1in}
\caption{Sketch completion rules.}
\label{fig:sketch-completion}
\vspace{-0.1in}
\end{figure}

\paragraph{Sketch completion algorithm.} Now that we can enumerate terms of type $\tau$, let us consider the full sketch completion procedure. Our algorithm is bottom-up 
and first synthesizes all arguments of type {\tt tbl} before synthesizing  other arguments. Given  sketch $\sketch$ and  example $\ex$, {\sc FillSketch}($\sketch, \ex$) returns a set of hypotheses representing \emph{complete programs}  that are valid with respect to our deduction system.

Our sketch completion procedure is described using the inference rules shown in Figure~\ref{fig:sketch-completion}.  The first rule corresponds to a  base case of the {\sc FillSketch} procedure and is used for completing hypotheses that are \emph{not} of type {\tt tbl}. Here, $\sketch$ represents a subpart of the sketch that we want to complete,  $\tab$ is the  table that should be used in completing $\sketch$,  and $\sketch_f$ is the full sketch. Since $\sketch$ represents an unknown expression of type $\tau_i$, we use the type inhabitation rules from Figure~\ref{fig:inhabit} to find a well-typed instantiation $t$ of $\tau_i$ with respect to table $\tab$. Given completion $t$ of $?_i$, the full sketch now becomes $\sketch_f[\sketch@t/\sketch]$, and we use the deduction system to check whether the new hypothesis is valid. Since our deduction procedure uses partial evaluation, we may now be able to obtain a concrete table for some part of the sketch, thereby enhancing the power of deductive reasoning. 
%Note that, if {\sc Deduce} returns $\bot$, we know that $t$ is not a valid instantiation of $?_i$; hence, we  discard it.

The second rule from Figure~\ref{fig:sketch-completion} is also a base case of the {\sc FillSketch} procedure. Since any leaf $?_i$ of type {\tt tbl} is already bound to some input variable $x$ in the sketch, there is nothing to complete; hence, we  just return $\sketch$ itself. 

Rule (3) corresponds to the recursive step of the {\sc FillSketch} procure and is used 
to complete a sketch with top-most component $\chi$. Specifically, consider a sketch of the form $?_i^\chi(\vec{\hypo}, \vec{\hypo'})$ where $\vec{\hypo}$ denotes arguments of type {\tt tbl} and $\vec{\hypo'}$ represents first-order functions. Since the vocabulary of $\vec{\hypo'}$ depends on the completion of $\vec{\hypo}$ (as explained earlier), we first recursively synthesize $\vec{\hypo}$ and obtain a set  of complete programs $\vec{\prog}$, together with their partial evaluation $\tab_1, \ldots, \tab_n$. Now, observe that each $\hypo_j' \in \vec{\hypo}'$ can refer to any of the columns in $\tab_1 \times ... \times \tab_n$; hence we recursively synthesize the remaining arguments $\vec{\hypo'}$ using table $\tab_1 \times ... 
\times \tab_n$. Now, suppose that the hypotheses $\vec{\hypo}$ and $\vec{\hypo'}$ are completed using terms $\vec{\prog}$ and $\vec{\prog}'$ respectively, and the new (partially filled) sketch is now $\sketch_f[\vec{\prog}/\vec{\hypo}, \vec{\prog'}/\vec{\hypo'}]$. Since there is  an opportunity for rejecting this partially filled sketch, we again check whether $\sketch_f[\vec{\prog}/\vec{\hypo}, \vec{\prog'}/\vec{\hypo'}]$ is consistent with the input-output examples using deduction.

\begin{figure}[t]
\begin{tabular}{ll}

     \begin{tabular}{c c c} 
     $\tab_3$  \\
     \hline
     \rowcolor{lightgoldenrodyellow}
     id & name & age \\ [0.5ex] 
     \hline\hline
     \rowcolor{babygreen}
     2 & Bob & 18  \\
     \hline
     \rowcolor{babypurple}
     3 & Tom & 12 \\
     \hline
    \end{tabular}
    &
         \begin{tabular}{c c c c} 
     $\tab_4$ \\
     \hline
    \rowcolor{lightgoldenrodyellow}
     id & name & age & GPA \\ [0.5ex] 
     \hline\hline
      \rowcolor{babygreen}
     2 & Bob & 18 & 3.2 \\
     \hline
    \end{tabular}
\end{tabular}
\caption{Tables for Example~\ref{ex:sketch}}\label{fig:sketch-tables}
\vspace{-0.1in}
\end{figure}

\begin{example}\label{ex:sketch}
Consider  hypothesis $\hypo$ from Figure~\ref{fig:tree},  the input table  $\tab_1$ from Figure~\ref{fig:partial-tables}, and the output table $\tab_3$ from Figure~\ref{fig:sketch-tables}. We can successfully
convert this hypothesis into the sketch $\lambda x. ?_0^\pi(?_1^\sigma(?_3@(x, \tab_1), ?_4), ?_2)$. % without obtaining a contradiction.
%\[
%\begin{array}{c}
%?_1.\rows < ?_3.\rows \ \land \  ?_1.\columns = ?_3.\columns \ \land \\?_0.\rows = ?_1.\rows \ \land \ ?_0.\columns < ?_1.\columns \land \\
%x_1 = ?_3 \land y = ?_0 \ \land x_1.\rows = 3 \land \\
%x_1.\columns = 4  \land   y.\rows = 2 \land y.\columns = 3 \\
%\end{array}
%\]
%Now consider completing this sketch using the {\sc FillSketch} procedure. 
Since {\sc FillSketch} is bottom-up,
it first tries to fill hole $?_4$. In this case, suppose that we try to instantiate hole 
$?_4$ with the predicate {\tt age > 12}  using rule (1) from Figure~\ref{fig:sketch-completion}. However, when we call {\sc Deduce} on the partially-completed sketch $\lambda x. ?_0^\pi(?_1^\sigma(?_3@(x, \tab_1), {\tt age}>12), ?_2)$, 
$?_1$ is refined as $\tab_4$ in Figure~\ref{fig:sketch-tables} and we obtain the following constraint:
\[
\begin{array}{c}
?_1.\rows < ?_3.\rows \ \land \  ?_1.\columns = ?_3.\columns \ \land ?_0.\rows = ?_1.\rows \ \land \\ \ ?_0.\columns < ?_1.\columns \land 
x_1 = ?_3  \ \land x_1.\rows = 3 \land x_1.\columns = 4 \ \land\\
 y = ?_0 \land   y.\rows = 2 \land y.\columns = 3 \land
  \text{\underline{\boldmath$?_1.\columns = 4  \ \land  \ ?_1.\rows = 1$}}
\end{array}
\]
Note that the last two conjuncts (underlined) are obtained using partial evaluation. Since this formula is unsatisfiable, we can reject this hypothesis without having to fill hole $?_2$.
\end{example}

\begin{figure*}[!t]
{\small 
\begin{center}
\begin{tabular}{|ccc|cc|cc|cc|}
\hline
\multirow{2}{*}{Category} & \multirow{2}{*}{Description} & \multirow{2}{*}{\#} & \multicolumn{2}{c|}{No deduction} & \multicolumn{2}{c|}{Spec 1} & \multicolumn{2}{c|}{Spec 2 }\\
&&&\#Solved & Time& \#Solved & Time& \#Solved & Time\\
\hline
\hline
C1 & \multicolumn{1}{m{5.6cm}}{\emph{Reshaping} dataframes from either ``long'' to ``wide'' or ``wide'' to ``long''} & 4 & 2 & 198.14 & 4 & 15.48 & 4 &6.70\\
\hline
C2 & \multicolumn{1}{m{5.6cm}}{\emph{Arithmetic computations} that produce values not present in the input tables} & 7 & 6 & 5.32 & 7 & 1.95 & 7 & 0.59\\
\hline
C3 & \multicolumn{1}{m{5.6cm}}{Combination of \emph{reshaping} and \emph{string manipulation} of cell contents} & 34 & 28 & 51.01 & 31 & 6.53 & 34 & 1.63\\
\hline
C4 & \multicolumn{1}{m{5.6cm}}{\emph{Reshaping} and \emph{arithmetic computations}} & 14 & 9 & 162.02 & 10 & 90.33 & 12 & 15.35\\
\hline
C5 & \multicolumn{1}{m{5.6cm}}{Combination of \emph{arithmetic computations} and \emph{consolidation} of information from multiple tables into a single table} & 11 & 7 & 8.72 & 10 & 3.16 & 11 & 3.17\\
\hline
C6 & \multicolumn{1}{m{5.6cm}}{\emph{Arithmetic computations} and \emph{string manipulation} tasks} & 2 & 1 & 280.61 & 2 & 49.33 & 2 & 3.03\\
\hline
C7 & \multicolumn{1}{m{5.6cm}}{\emph{Reshaping} and \emph{consolidation} tasks} & 1 & 0 & \xmark & 1 & 135.32 & 1 & 130.92\\
\hline
C8 & \multicolumn{1}{m{5.6cm}}{Combination of \emph{reshaping}, \emph{arithmetic computations} and \emph{string manipulation}} & 6 & 1 & \xmark & 3 & 198.42 & 6 & 38.42\\
\hline
C9 & \multicolumn{1}{m{5.6cm}}{Combination of \emph{reshaping}, \emph{arithmetic computations} and  \emph{consolidation}} & 1 & 0 & \xmark & 0 & \xmark & 1 & 97.3\\
\hline
\hline
\multicolumn{2}{|c}{\multirow{2}{*}{Total}} & \multirow{2}{*}{80} & 54 & \multirow{2}{*}{95.53} & 68 & \multirow{2}{*}{8.57} & 78 & \multirow{2}{*}{3.59}\\
&&&(67.5\%) && (85.0\%) && (97.5\%)&\\
\hline
\end{tabular}
\end{center}
}
\vspace{-0.1in}
\caption{Summary of experimental results. All times are median in seconds and \xmark \ indicates a timeout ($>$ 5 minutes).}
%\caption{\toolname performance using multiple specifications with different levels of precision}
 %Properties P1, P2 and P3 correspond to reshaping, computation and consolidation tasks. Times are in seconds and $\bot$ indicates a timeout ($>$ 5 minutes).}
 \label{tbl:results}
\vspace{-0.1in}
\end{figure*}

\section{Implementation}\label{sec:impl}
We have implemented our synthesis algorithm in a tool called
\toolname, written in C++. 
%which consists of approximately 5,600 lines of C++ code.
\toolname uses the Z3 SMT solver~\cite{z3} with the theory of Linear Integer Arithmetic 
for checking the satisfiability of constraints generated by our deduction engine. 

Recall from Section~\ref{sec:alg} that \toolname uses a cost model for
picking the ``best" hypothesis from the worklist. Inspired by previous
work on  code completion~\cite{raychev2014}, we use a cost model based
on a statistical analysis of existing code. Specifically, \toolname analyzes existing code snippets that use components from $\comps_\tab$ and  represents each  snippet as
 a `sentence' where  `words' correspond to  components in $\comps_\tab$. Given this representation, \toolname uses the 2-gram model in SRILM~\cite{SRILM} to assign a score to each hypothesis. The  hypotheses in 
the worklist $\worklist$ from Algorithm~\ref{alg:synthesis} are then ordered using the scores obtained from the $n$-gram model.

\begin{comment}
In addition to the core algorithms described in previous sections
our implementation performs two main optimizations. First, we use 
an n-gram model to rank all possible candidate hypotheses $\hypo$. 
Specifically, \toolname adopts the 2-gram model in SRILM~\cite{SRILM}. 
To obtain the training data, we crawled around 15,000 code snippets from 
Stackoverflow using the keywords {\tt tidyr} and {\tt dplyr}. 
For each code snippet, we ignore its control flow and represent it  
using a `sentence' where each `word' corresponds to an API call.
Finally, we use the n-gram model to rank the candidate hypotheses in 
the worklist $\worklist$ mentioned in Algorithm~\ref{alg:synthesis}.
\end{comment}

Following the \emph{Occam's razor} principle, \toolname explores hypotheses in increasing order of size.
%
%size $1,2,...,n$ sequentially. 
However, if the size of the correct hypothesis is a large number $k$, \toolname may end up exploring many programs before reaching length $k$. In practice, we have found that a better strategy is to 
exploit the inherent parallelism of our algorithm.
% and use multiple threads to 
%simultaneously explore hypotheses of different sizes. 
Specifically, \toolname uses 
multiple threads to search for solutions of different sizes and terminates as soon as any  
thread finds a correct solution.

%To battle against
%this issue, the second optimization is using multiple threads to explore 
%hypotheses of different size simultaneously and \toolname terminates as soon
%as any thread finds the correct solution. In practice we use one thread to 
%handle hypotheses from size $1$ to $4$ and another thread to handle hypotheses
%that are greater than 4.

\section{Evaluation}\label{sec:eval}

To evaluate our method, we collected 80 data preparation tasks, all of which are drawn from discussions among R users on Stackoverflow.
The supplementary material contains (i) the Stackoverflow post for each benchmark, (ii)  
an input-output example, and (iii) the solution synthesized by \toolname.

%The supplementary material contains (i) a detailed 
%description of each benchmark, (ii) the corresponding Stackoverflow post, (iii)  
%an input-output example, and (iv) the solution synthesized by \toolname.
%
Our evaluation aims to answer the following 
questions:

\begin{itemize}
\item[{\bf Q1.}] Can \toolname successfully automate real-world data preparation 
tasks and what is its running time?

\item[{\bf Q2.}] How big are the benefits of SMT-based deduction and partial evaluation in the performance of \toolname?

\item[{\bf Q3.}] How complex are the data preparation tasks that can be successfully automated using  \toolname?

\item[{\bf Q4.}] Are there existing synthesis tools that can also automate the data preparation tasks supported by \toolname?
\end{itemize}

To answer these questions, we performed a series of experiments on the 80 data preparation benchmarks, using the input-output examples provided by the authors of the Stackoverflow posts. In these experiments, we use ten table transformation components from {\tt tidyr} and {\tt dplyr}, two popular table manipulation libraries for R. In addition, we also use ten value transformation components, including the standard comparison operators such as {\tt <} , {\tt >}  as well as aggregate functions like {\tt MEAN} and {\tt SUM}. 
All experiments are conducted
on an Intel Xeon(R) computer with an E5-2640 v3 CPU and 32G
of memory, running the Ubuntu 14.04 operating system and using a timeout of 5 minutes.

\paragraph{{\bf Summary of results.}} The results of our evaluation are 
summarized in Figure~\ref{tbl:results}. Here, the \emph{``Description''} column provides a brief English description of each category, and the column ``\#" shows the number of benchmarks in each category. The \emph{``No deduction''} column indicates the running time of a version of \toolname that uses purely enumerative search without deduction. (This  basic version still uses the statistical analysis described in Section~\ref{sec:impl} to choose the ``best'' hypothesis.)  The columns labeled \emph{``Spec 1''} and \emph{``Spec 2''} show  variants of \toolname using two different component specifications. Specifically, \emph{Spec 1} is less precise and only constrains the relationship between the number of rows and columns, as shown in Table~\ref{table:spec}. On the other hand, \emph{Spec 2} is strictly more precise than \emph{Spec 1} and also uses other information, such as cardinality and number of groups. 
%\todo{The interested reader can find these two different specifications  under supplementary materials.}

%inspected the synthesized code and observed that even with a single example, 
%\toolname was able to synthesize the correct R implementation for all 
%benchmarks. 
%Note that if the input-output example is too small then it may be 
%the case that \toolname would find an alternative solution that does not 
%reflect user intent. Even though this was not the case in our benchmarks, if
%this scenario would happen we could increase the size of the examples and 
%re-run the tool. 
%

\begin{comment}
Table~\ref{tbl:results} shows the performance of \toolname when using multiple 
specifications with different levels of precision. ``Spec 1'' is the simplest
specification that only asserts true and does not exploit any properties of the 
table. ``Spec 2'' uses specifications that describe the number of columns 
and rows of the input tables and its corresponding output table when a specific 
high-order component is used. ``Spec 3'' uses cardinality based specifications 
in addition to the number of columns and rows used in Spec 2. Each specification 
has 10 high-order components and their detailed specifications can be found in the 
supplementary material. For Specs 2 and 3 we also present the results with (w/ p.e.) 
and without partial evaluation (w/o p.e.).
\end{comment}

\paragraph{{\bf Performance.}} As shown in Figure~\ref{tbl:results}, the full-fledged version of \toolname (using the more precise component specifications) can successfully synthesize 78 out of the 80 benchmarks and times out on only 2 problems. Hence, overall, \toolname achieves a success rate of 97.5\% within a 5-minute time limit.  \toolname's median running time on these benchmarks is 3.59 seconds, and 86.3\% of the benchmarks can be synthesized within 60 seconds. However, it is worth noting that running time is actually dominated by the R interpreter: \toolname spends roughly 68\% of the time in the R interpreter, while  using only 15\% of its running time to perform deduction (i.e., solve SMT formulas). Since the overhead of the R interpreter can be significantly reduced with sufficient engineering effort, we believe there is considerable room for improving \toolname's running time. However, even in its current form, these results show that \toolname is practical enough to automate a diverse class of data preparation tasks within a reasonable time limit.

\paragraph{{\bf Impact of deduction.}} As Figure~\ref{tbl:results} shows, deduction has a huge positive impact on the algorithm. The basic version of \toolname that does not perform deduction times out on 32.5\% of the benchmarks and achieves a median running time of 95.53 seconds. On the other hand,  if we use the coarse specifications given by \emph{Spec 1}, we already observe a significant improvement. Specifically, using \emph{Spec 1}, \toolname can successfully solve 68 out of the 80 benchmarks, with a median running time of 8.57 seconds. These results show that even coarse and easy-to-write specifications can have a significant positive impact on synthesis.

\paragraph{{\bf Impact of partial evaluation.}} 
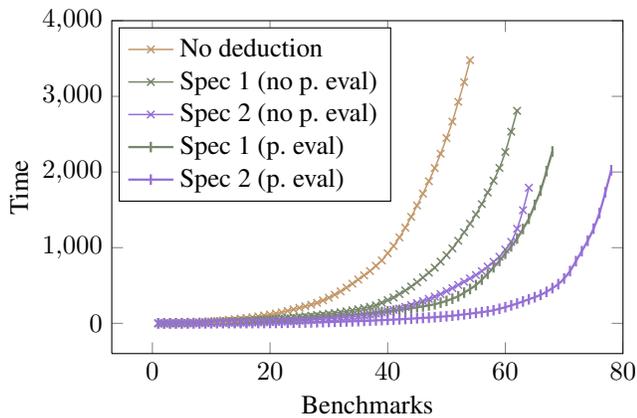
\begin{figure}
\begin{tikzpicture}
     \begin{axis}[
     ymax=4000,
     y=0.001cm,
     x=2.225,
    legend cell align=left,
    legend pos = outer north east,
    legend style={
        at={(0.28,0.98)},
        legend columns=1,
        anchor=north
    },
    xlabel style={yshift=1mm},
    ylabel = Time,
    xlabel = Benchmarks,
    xmax = 80,
     ]
     \legend{No deduction,Spec 1 (no p. eval),Spec 2 (no p. eval),Spec 1 (p. eval),Spec 2 (p. eval)} 
      \addplot[
        smooth,
        line width=0.2mm,
         mark=x,
         camel
       ] plot coordinates {
     (  1 , 0.04  )
( 2 , 0.18  )
( 3 , 0.66  )
( 4 , 1.28  )
( 5 , 2.16  )
( 6 , 4.25  )
( 7 , 6.94  )
( 8 , 11.84 )
( 9 , 17.16 )
( 10  , 24.45 )
( 11  , 31.92 )
( 12  , 39.47 )
( 13  , 47.08 )
( 14  , 54.72 )
( 15  , 62.72 )
( 16  , 70.93 )
( 17  , 79.65 )
( 18  , 91.52 )
( 19  , 103.89  )
( 20  , 116.52  )
( 21  , 129.35  )
( 22  , 145.5 )
( 23  , 163.83  )
( 24  , 183.88  )
( 25  , 204.03  )
( 26  , 224.55  )
( 27  , 246.57  )
( 28  , 270.76  )
( 29  , 303.47  )
( 30  , 337.93  )
( 31  , 378.06  )
( 32  , 422.02  )
( 33  , 472.11  )
( 34  , 524.04  )
( 35  , 578.61  )
( 36  , 633.86  )
( 37  , 691.05  )
( 38  , 760.21  )
( 39  , 831.08  )
( 40  , 925.85  )
( 41  , 1022.13 )
( 42  , 1135.53 )
( 43  , 1264.07 )
( 44  , 1407.32 )
( 45  , 1556.84 )
( 46  , 1714.19 )
( 47  , 1878.16 )
( 48  , 2052.68 )
( 49  , 2241.38 )
( 50  , 2448.64 )
( 51  , 2666.69 )
( 52  , 2927.4  )
( 53  , 3188.61 )
( 54  , 3477.31 )
       };
      \addplot[
        smooth,
        line width=0.2mm,
         mark=x,
         camouflagegreen
       ] plot coordinates {
( 1 , 0.03  )
( 2 , 0.14  )
( 3 , 0.39  )
( 4 , 0.84  )
( 5 , 1.37  )
( 6 , 1.96  )
( 7 , 2.78  )
( 8 , 4.01  )
( 9 , 5.63  )
( 10  , 7.29  )
( 11  , 9.59  )
( 12  , 13.38 )
( 13  , 17.61 )
( 14  , 21.84 )
( 15  , 26.56 )
( 16  , 31.6  )
( 17  , 36.8  )
( 18  , 42.22 )
( 19  , 47.68 )
( 20  , 53.37 )
( 21  , 59.18 )
( 22  , 65.37 )
( 23  , 71.93 )
( 24  , 78.99 )
( 25  , 86.27 )
( 26  , 93.61 )
( 27  , 101.23  )
( 28  , 109.44  )
( 29  , 118.05  )
( 30  , 126.66  )
( 31  , 135.91  )
( 32  , 145.58  )
( 33  , 158.35  )
( 34  , 171.76  )
( 35  , 186.2 )
( 36  , 201.19  )
( 37  , 217.89  )
( 38  , 237.78  )
( 39  , 271.79  )
( 40  , 306.12  )
( 41  , 341.28  )
( 42  , 386.98  )
( 43  , 433.87  )
( 44  , 486.11  )
( 45  , 543.2 )
( 46  , 605.31  )
( 47  , 669.95  )
( 48  , 735.49  )
( 49  , 815.43  )
( 50  , 901.39  )
( 51  , 990.44  )
( 52  , 1091.49 )
( 53  , 1202.95 )
( 54  , 1316.79 )
( 55  , 1439.59 )
( 56  , 1575.26 )
( 57  , 1729.46 )
( 58  , 1885.86 )
( 59  , 2053.61 )
( 60  , 2263.7  )
( 61  , 2533.1  )
( 62  , 2808.97 )
       };
             \addplot[
        smooth,
        line width=0.2mm,
         mark=x,
         darkpastelpurple
       ] plot coordinates {
( 1 , 0.04  )
( 2 , 0.15  )
( 3 , 0.32  )
( 4 , 0.67  )
( 5 , 1.05  )
( 6 , 1.48  )
( 7 , 1.96  )
( 8 , 2.52  )
( 9 , 3.33  )
( 10  , 4.2 )
( 11  , 5.08  )
( 12  , 5.98  )
( 13  , 6.94  )
( 14  , 7.92  )
( 15  , 8.94  )
( 16  , 10.11 )
( 17  , 11.42 )
( 18  , 13.2  )
( 19  , 15.05 )
( 20  , 16.96 )
( 21  , 19.09 )
( 22  , 21.91 )
( 23  , 25.35 )
( 24  , 29.19 )
( 25  , 33.43 )
( 26  , 38.01 )
( 27  , 43.31 )
( 28  , 48.66 )
( 29  , 54.48 )
( 30  , 60.32 )
( 31  , 66.25 )
( 32  , 72.61 )
( 33  , 79.17 )
( 34  , 86.08 )
( 35  , 93.18 )
( 36  , 103.3 )
( 37  , 115.21  )
( 38  , 128.6 )
( 39  , 143.54  )
( 40  , 160.45  )
( 41  , 177.67  )
( 42  , 195.14  )
( 43  , 216.92  )
( 44  , 241.84  )
( 45  , 266.97  )
( 46  , 294.21  )
( 47  , 322.61  )
( 48  , 353 )
( 49  , 386.11  )
( 50  , 425.26  )
( 51  , 465.29  )
( 52  , 507.03  )
( 53  , 549.47  )
( 54  , 593.54  )
( 55  , 637.8 )
( 56  , 690.16  )
( 57  , 750.02  )
( 58  , 812.47  )
( 59  , 880.52  )
( 60  , 975.24  )
( 61  , 1076.95 )
( 62  , 1250.51 )
( 63  , 1494.85 )
( 64  , 1791.73 )
       };
             \addplot[
        smooth,
        line width=0.3mm,
         mark=|,
         camouflagegreen
       ] plot coordinates {
( 1 , 0.04  )
( 2 , 0.14  )
( 3 , 0.38  )
( 4 , 0.85  )
( 5 , 1.35  )
( 6 , 1.85  )
( 7 , 2.67  )
( 8 , 3.97  )
( 9 , 5.64  )
( 10  , 7.32  )
( 11  , 9.26  )
( 12  , 11.21 )
( 13  , 13.72 )
( 14  , 16.3  )
( 15  , 19.44 )
( 16  , 22.6  )
( 17  , 26  )
( 18  , 29.45 )
( 19  , 32.9  )
( 20  , 36.66 )
( 21  , 40.43 )
( 22  , 44.29 )
( 23  , 48.24 )
( 24  , 53.12 )
( 25  , 58.05 )
( 26  , 62.98 )
( 27  , 68.45 )
( 28  , 74.17 )
( 29  , 79.98 )
( 30  , 85.91 )
( 31  , 92.03 )
( 32  , 98.24 )
( 33  , 105.18  )
( 34  , 112.28  )
( 35  , 119.56  )
( 36  , 126.91  )
( 37  , 134.47  )
( 38  , 142.21  )
( 39  , 149.95  )
( 40  , 158.28  )
( 41  , 167.09  )
( 42  , 176.19  )
( 43  , 185.98  )
( 44  , 198.25  )
( 45  , 210.83  )
( 46  , 223.7 )
( 47  , 236.96  )
( 48  , 252 )
( 49  , 272.72  )
( 50  , 296.4 )
( 51  , 325.85  )
( 52  , 357.78  )
( 53  , 405.04  )
( 54  , 453.2 )
( 55  , 504.59  )
( 56  , 577.11  )
( 57  , 655.05  )
( 58  , 740.17  )
( 59  , 825.33  )
( 60  , 920.05  )
( 61  , 1016.89 )
( 62  , 1119.61 )
( 63  , 1247.31 )
( 64  , 1382.63 )
( 65  , 1565.34 )
( 66  , 1755.67 )
( 67  , 2010.87 )
( 68  , 2271.19 )
       };
       \addplot[
       line width=0.3mm,
        smooth,
         mark=|,
         darkpastelpurple
       ] plot coordinates {
( 1 , 0.07  )
( 2 , 0.14  )
( 3 , 0.27  )
( 4 , 0.41  )
( 5 , 0.57  )
( 6 , 0.74  )
( 7 , 0.91  )
( 8 , 1.1 )
( 9 , 1.31  )
( 10  , 1.52  )
( 11  , 1.74  )
( 12  , 1.96  )
( 13  , 2.19  )
( 14  , 2.43  )
( 15  , 2.69  )
( 16  , 2.96  )
( 17  , 3.25  )
( 18  , 3.58  )
( 19  , 4.07  )
( 20  , 4.63  )
( 21  , 5.22  )
( 22  , 6.16  )
( 23  , 7.16  )
( 24  , 8.22  )
( 25  , 9.57  )
( 26  , 11.03 )
( 27  , 12.55 )
( 28  , 14.09 )
( 29  , 15.82 )
( 30  , 17.85 )
( 31  , 19.95 )
( 32  , 22.13 )
( 33  , 24.4  )
( 34  , 26.71 )
( 35  , 29.36 )
( 36  , 32.12 )
( 37  , 35.29 )
( 38  , 38.69 )
( 39  , 42.12 )
( 40  , 45.67 )
( 41  , 49.3  )
( 42  , 52.94 )
( 43  , 57.23 )
( 44  , 61.78 )
( 45  , 66.35 )
( 46  , 71.24 )
( 47  , 76.25 )
( 48  , 81.28 )
( 49  , 87.41 )
( 50  , 93.66 )
( 51  , 100.81  )
( 52  , 109.4 )
( 53  , 118.45  )
( 54  , 127.66  )
( 55  , 137.62  )
( 56  , 147.82  )
( 57  , 158.47  )
( 58  , 172.76  )
( 59  , 191.91  )
( 60  , 213.4 )
( 61  , 236.83  )
( 62  , 261.62  )
( 63  , 289.15  )
( 64  , 317.09  )
( 65  , 345.18  )
( 66  , 376.81  )
( 67  , 416.12  )
( 68  , 465.01  )
( 69  , 518.13  )
( 70  , 593.27  )
( 71  , 690.57  )
( 72  , 821.49  )
( 73  , 953.34  )
( 74  , 1091.34 )
( 75  , 1251.66 )
( 76  , 1456.49 )
( 77  , 1731.64 )
( 78  , 2023.6  )
       };
     \end{axis}
   \end{tikzpicture}
   \vspace{-0.2in}
   \caption{Cumulative running time of \toolname}\label{fig:pe}
   \vspace{-0.1in}
\end{figure}

%Partial evaluation plays a major role in the performance of \toolname. 
Figure~\ref{fig:pe} shows the cumulative running time of \toolname with and without partial evaluation. 
Partial evaluation significantly improves the performance of \toolname,  both in terms of running time and the number of benchmarks solved.
In particular, without partial evaluation, \toolname can only solve 62 benchmarks with median running time of 34.75 seconds using \emph{Spec 1} 
and 64 benchmarks with median running time of 17.07 seconds using \emph{Spec 2}.
%
%with median running times of 45.29 and 33.89 seconds using \emph{Spec 1} and \emph{Spec 2}, respectively. 
%When partial evaluation is used, \toolname (with \emph{Spec 1}) can solve 69 benchmarks with a median running time of 8.88 seconds and, using \emph{Spec 2 can solve 78 benchmarks with a median running time of 6.09 seconds. 
When using partial evaluation, \toolname can prune 72\% of the partial programs without having to fill all holes in the sketch, thereby resulting in significant performance improvement.

\paragraph{{\bf Complexity of benchmarks.}} To evaluate the complexity of tasks that \toolname can handle, we conducted a small user study involving 9 participants. Of the participants,  four are senior software engineers at a leading data analytics company and do data preparation ``for a living". The remaining 5 participants are proficient R programmers at a university and specialize in statistics, business analytics, and machine learning. We chose 5 representative examples from our 80 benchmarks and 
asked the participants to solve as many of them as possible within one hour. 
These benchmarks belong to four categories (C2, C3, C4, C7) and take between 0.22 and 204.83 seconds to be solved by \toolname.%, with a total solving time of 348.08 seconds for all five.

%\todo{Need to say something about these 5 benchmarks and why they are actually representative.}

In our  user study, the average participant completed $3$ tasks within the one-hour time limit; however, only 2 of these tasks were solved \emph{correctly} on average. These results suggest that our benchmarks are challenging even for proficient R programmers and expert data analysts. %Hence, we believe that tools like \toolname have great potential to improve the productivity of data analysts in real-world data preparation tasks.

\paragraph{{\bf Comparison with other tools.}} 

\begin{figure}[!t]
\centering
\begin{tikzpicture}
%\small
    \begin{axis}[
        width  = 8cm,
        height = 7.5cm,
        y=0.026cm,
        x=1.4cm,
        major x tick style = transparent,
        ybar=2*\pgflinewidth,
        bar width=11pt,
        ymajorgrids = true,
        ylabel = {Percentage},
        ylabel style={yshift=-4mm},
        xlabel style={yshift=3mm},
        symbolic x coords={R,SQL},
        xtick = {data},
        scaled y ticks = false,
        enlarge x limits=0.8,
        ymin=0,
        ymax=100,
        legend cell align=left,
        legend pos = outer north east,
        legend style={
                at={(0.4,1.3)},
                legend columns=-1,
                anchor=north,
        %        column sep=1ex
        }
    ]
    \hspace*{-2mm}
        \addplot[style={camouflagegreen,fill=camouflagegreen,mark=none}]
            %coordinates {(R,1)(SQL,19)};
            coordinates {(R,1.3)(SQL,71.4)};

        \addplot[style={babypurple,fill=babypurple,mark=none}]
             %coordinates {(R,79)(SQL,27)};
             coordinates {(R,98.8)(SQL,96.4)};

        \legend{\sql,\toolname}
    \end{axis}
\end{tikzpicture}
\vspace{-0.05in}
\caption{Comparison with \sql}
\vspace{-0.1in}
\label{fig:tools}
\end{figure}

To demonstrate the advantages of our proposed approach over previous techniques, we compared \toolname 
with $\lambda^2$~\cite{lambda2} and \sql~\cite{zhang-ase2013}. Among these, $\lambda^2$ is a fairly general approach for synthesizing higher-order functional programs over data structures. In contrast, \sql is a more specialized tool for synthesizing SQL queries from examples.

Since $\lambda^2$ does not have built-in support for tables, we evaluated $\lambda^2$ on the benchmarks from Figure~\ref{tbl:results} by representing each table as a list of lists. Even though we confirmed that $\lambda^2$ can synthesize very simple table transformations involve projection and  selection, it was not able to successfully synthesize \emph{any} of the benchmarks used in our evaluation.

\begin{comment}
$\lambda^2$ is a tool for synthesizing higher-order 
functional programs 
%from input-output examples but can also 
that can synthesize table 
transformations by viewing tables as list of lists. Though $\lambda^2$ 
can synthesize simple table transformations that either involve projection and 
selection in relational algebra or aggregate functions, it cannot synthesize 
more complex transformations that combine these operators. We 
ran $\lambda^2$ on the 6 benchmarks from category C2. Even though these benchmarks involve fewer 
combinations of operators, 
$\lambda^2$ was not able 
to synthesize any of them.
\end{comment}

To compare \toolname with \sql, we used two different sets of benchmarks. First, we evaluated \sql on the 80 data preparation benchmarks from Figure~\ref{tbl:results}. Note that some of the data preparation tasks used in our evaluation cannot be expressed using SQL, and therefore fall beyond the scope of a tool like \sql. Among our 80 benchmarks, \sql was only able to successfully solve \emph{one}.

%\sql is a tool for synthesizing SQL queries from examples. 
%Of the 80 R benchmarks, \sql can only solve 2.
%As R is more expressive than SQL, some benchmarks involve complex table 
%transformations that are not supported by SQL. For the benchmarks supported by 
%SQL, \sql can only solve a subset of them since this tool only supports a 
%fragment of SQL.
%

%As \toolname can work with any set of components, it 
%may be used in domains other than R. To illustrate this, 
To understand how \toolname compares with \sql on a narrower set of table transformation tasks, we also evaluated both tools on the 28 benchmarks used in evaluating \sql~\cite{zhang-ase2013}. To solve these benchmarks using \toolname, we 
used the same input-output tables as \sql and used a total of eight higher-order components that are relevant to SQL. As shown in Figure~\ref{fig:tools}, \toolname also outperforms \sql on these benchmarks. In particular, \toolname can solve 96.4\% of the SQL benchmarks with a median running time of 1 second whereas \sql can solve only 71.4\% with a median running time of 11 seconds.

\section{Related Work}

In this section, we relate our approach to prior work on synthesis and techniques for facilitating data transformations.

%Program synthesis has received significant attention in the programming
%languages community in the last several years.  Many approaches to the
%problem require a programmer-provided ``template'' or partial program
%~\cite{sketch-asplos06,SrivastavaGF13,BeyeneCPR14} or a complete logical
%specification~\cite{completefunctional}. In contrast, our method
%performs synthesis from examples.
\vspace{0.05in}
\noindent
{\bf \emph{PBE for table transformations.}} This paper is related to a line of work on programming-by-example (PBE)~\cite{example-popl11,example-pldi11,flashrelate,comp-pldi14,kitzelmann11,example-cav13,lambda2,myth,flashextract,navid-pldi16,flashmeta}. Of
particular relevance are PBE techniques that focus on table
transformations~\cite{example-pldi11,zhang-ase2013,flashextract,flashrelate}. Among
these techniques, {\sc FlashExtract} and {\sc FlashRelate} address the
specific problem of extracting structured data from spreadsheets and
do not consider a general class of table transformations. More closely
related are Harris and Gulwani's work on synthesis of 
spreadsheet transformations~\cite{example-pldi11} and Zhang et
al.'s work on synthesizing SQL
queries~\cite{zhang-ase2013}. 
Our approach is more general than these methods in that they 
use DSLs with a fixed set of
primitive operations (components), whereas our approach takes a set of
components as a {\em parameter}. 
For instance,  Zhang et al. cannot synthesize programs that perform
table reshaping while Harris et al. supports data reshaping, but not
computation or consolidation. Hence, these approaches cannot 
automate many of the data preparation tasks that we consider.

\vspace{0.05in}
\noindent
{\bf \emph{Data wrangling.}} Another  term for data preparation is
\emph{``data wrangling"}, and 
prior work has considered methods to facilitate such tasks. For instance, {\sc Wrangler} is an interactive visual system that aims to simplify data wrangling~\cite{wrangler1,wrangler2}. {\sc OpenRefine} is a general framework that helps users perform data transformations and clean messy data. % and parse data from websites~\cite{openrefine}. 
Tools such as {\sc Wrangler} and {\sc OpenRefine} facilitate a larger class of data wrangling tasks than {\sc Morpheus}, but they do not automatically synthesize table transformations from  examples.

%Also, Zhang et al. do not consider
%synthesis queries with nested table operations. There are, however,
%some facets on which the approaches are incomparable. In particular,
%our approach uses the fact that tables are {\em typed}, whereas the
%spreadsheets considered by Harris and Gulwani~\cite{example-pldi11}
%are untyped (e.g., the same column can contain both numbers and
%strings).

\vspace{0.05in}
\noindent
{\bf \emph{Synthesis using deduction and search.}} 
Our work builds on  recent synthesis techniques that combine
enumeration and
deduction~\cite{lambda2,myth,flashextract,navid-pldi16,example-cav13}. The closest work
in this space is $\lambda^2$, which  synthesizes functional programs
using deduction and cost-directed
enumeration~\cite{lambda2}. Like $\lambda^2$, we differentiate between higher-order and first-order combinators and use deduction to prune partial programs. However, the key  difference from prior techniques is that our
deduction capabilities are not customized to a specific set of components. 
For example, $\lambda^2$ only supports a fixed set of higher-order combinators and uses ``baked-in" deductive reasoning to reject partial programs.  In contrast, our approach  supports any higher-order component and can utilize arbitrary first-order specifications to reject hypotheses using SMT solving.

%uses SMT-based reasoning to exploit the 

%can efficiently search through a space of
%compositions of components using SMT-based
%deduction and  partial evaluation.
%Specifically, {\sc FlashExtract} synthesizes programs made from higher-order
%combinators using a custom deductive procedure~\cite{flashextract},
%and Escher uses goal-directed enumerative search to synthesize
%first-order programs with recursive functions~\cite{example-cav13}.
%The $\lambda^2$ system synthesizes higher-order functional programs
%using a combination of deduction and cost-directed
%enumeration~\cite{lambda2}.
% and Yaghmazadeh et al.~\cite{navid-pldi16}
%synthesize tree transformations through a reduction to the problem of
%synthesizing  path transformations.
%The central difference between our work and these efforts is that our
%deduction capabilities are not customized to a specific set of components. 
%In particular, given any set of components with a corresponding first-order specification, 
%our approach  can efficiently search through a space of
%compositions of components using SMT-based
%deduction and  partial evaluation.
%which are not used in
%prior efforts.

Also related  is {\sc FlashMeta}, which gives a generic  
method for constructing example-driven synthesizers for user-defined
DSLs~\cite{flashmeta}.  The methodology we propose in this paper is quite different from {\sc FlashMeta}. {\sc FlashMeta} uses version space algebras to represent \emph{all} programs consistent with the examples and employs deduction to decompose the synthesis task. In contrast, we use enumerative search to find \emph{one} program that satisfies the examples and use SMT-based deduction to reject partial programs.

%for synthesis methods based on search
%and deduction. Our synthesis problem could in principle be expressed
%in the language framework of this work. However, the central
%algorithmic ideas of our work, for example the use of partial
%evaluation and the use of SMT-based pruning, are not discussed in that
%paper.

%Also, Le and
%Gulwani~\cite{gulwani-flash-extract} present a synthesis framework for
%DSLs where programs are made from combinators such as map and
%filter. The problem domains targeted by these approaches and ours are
%similar; in fact, our synthesis problem could in principle be
%expressed in the DSL framework of the former work. However,
%the synthesis algorithms in these approaches are based purely on
%enumerative search, and are therefore unlikely to perform well on our
%experimental benchmarks.

\vspace{0.05in}
\noindent
{\bf \emph{Component-based synthesis.}}
Component-based synthesis  refers to generating (straight-line) programs from a  set of components, such as methods provided by an API~\cite{comp-pldi11,comp-icse10,api-pldi05,JohnsonE06,sypet}. Some of these efforts~\cite{comp-icse10,comp-pldi11} use an SMT-solver to {\em
  search} for a composition of components. In contrast, our approach
uses an SMT-solver as a {\em pruning tool} in enumerative search and does not require precise specifications of components. Another related work in this space is {\sc SyPet}~\cite{sypet}, which searches for well-typed programs using a Petri net representation. Similar to this work, {\sc SyPet} can also work with any set of components and decomposes synthesis into two separate sketch generation and sketch completion phases. However, both the application domains (Java APIs vs. table transformations) and the underlying techniques (Petri net reachability vs. SMT-based deduction) are very different.

\vspace{0.05in}
\noindent
{\bf \emph{Synthesis as type inhabitation.}}
Our approach views sketch completion as a type
inhabitation problem. In this respect, it resembles  prior
work that has framed synthesis
as type inhabitation~\cite{GveroKKP13,myth,FrankleOWZ16,synquid}. 
Of these approaches, {\sc InSynth}~\cite{GveroKKP13}
is type-directed rather than example-directed. {\sc Myth}~\cite{myth}
and its successors~\cite{FrankleOWZ16} cast type- and example-directed
synthesis as type inhabitation in a refinement type system. {\sc Synquid}~\cite{synquid} steps 
further by taking advantage of recent advances in polymorphic refinement types~\cite{liquidtype,ranjit16}. In contrast to these techniques,
our approach only enumerates type inhabitants in the context of sketch completion and uses table contents
 to finitize the universe of type inhabitants.

\vspace{0.05in}
\noindent
{\bf \emph{Sketch.}} In \emph{program sketching}, the user provides a partial program containing holes, which are completed by the synthesizer in a way that respects user-provided invariants (e.g., assertions)~\cite{sketch-asplos06,sketch-pldi05,sketch-pldi07}.  While we also use the  term \emph{``sketch"} to denote partial programs with unknown expressions, the holes in our program sketches can be arbitrary expressions over first-order components. In contrast, holes in the {\sc Sketch} system typically correspond to constants~\cite{sketch-pldi07}. Furthermore, our approach automatically generates program sketches rather than requiring the user to provide the sketch.
%While our
%problem statement can be seen as an instance of the foundational
%framework proposed in \cite{FrankleOWZ16}, that work does not offer
%concrete algorithmic strategies such as the use of SMT solvers or
%^partial evaluation, which make a huge difference to the 
%performance of our system.

\section{Conclusion}

We have presented a new synthesis algorithm for automating a large class of table transformation tasks that commonly arise in data science. Since our approach can work with any set of combinators and their corresponding specification, our synthesis algorithm is quite flexible and achieves scalability using SMT-based deduction and partial evaluation. As shown in our experimental evaluation, our tool, \toolname, can automate challenging data preparation tasks that are difficult even for proficient R programmers.

\bibliographystyle{abbrvnat}
{\small
\bibliography{main}
}

\newcommand*{\appendixes}{} %comment to remove comments
\ifdefined\appendixes
%\newpage
\appendix
\section*{Appendix A: Specifications of high-order components}

In this section, we present two specifications used in 
Section~\ref{sec:eval}. Specifically, as it is shown in 
table~\ref{table:spec1}, \emph{Spec 1} only constrains the relationship 
between the number of rows and columns. For instance, $\tab.\columns$ 
represents the number of columns and $\tab.\rows$ 
represents the number of rows of table $\tab$.

%For instance, $\tab_{out}.\columns$ 
%represents the number of columns of the output table.

% In this section, we show two different specifications which are used
% in the experiment. Specifically, as it is shown in table~\ref{table:spec1}, \emph{Spec 1} 
% only constrains the relationship between the number of
% rows and columns. For instance, $\tab_{out}.\columns$ represents the number of
% columns of the output table.

\begin{table*}
\vspace{-0.1in}
\centering
\newcolumntype{A}{ >{\arraybackslash} m{0.41\textwidth} }
\newcolumntype{B}{ >{\centering\arraybackslash} m{0.085\textwidth} }
\newcolumntype{E}{ >{\centering\arraybackslash} m{0.24\textwidth} }
\newcolumntype{D}{ >{\centering\arraybackslash} m{0.03\textwidth} }
\scalebox{0.9}{
\begin{tabular}{| D | B | A | E |}
   \hline
    Lib & Component & \centering Description & Specification \\
\hline
    \multirow{8}{*}{\begin{turn} {90}\makecell{tidyr}\end{turn}}
&  spread & Spread a key-value pair across multiple columns. &  $\tab_{out}.\rows \leq \tab_{in}.\rows$ \newline $\tab_{out}.\columns \geq \tab_{in}.\columns$ \\
\cline{2-4}
&  gather & Takes multiple columns and collapses into key-value pairs, duplicating all other columns as needed.& $\tab_{out}.\rows \geq \tab_{in}.\rows$ \newline $\tab_{out}.\columns \leq \tab_{in}.\columns$ \\
\cline{2-4}
&  separate & Separate one column into multiple columns.& $\tab_{out}.\rows = \tab_{in}.\rows$ \newline $\tab_{out}.\columns = \tab_{in}.\columns + 1$ \\
\cline{2-4}
&  unite & Unite multiple columns into one.& $\tab_{out}.\rows = \tab_{in}.\rows$ \newline $\tab_{out}.\columns = \tab_{in}.\columns - 1$ \\
\cline{2-4}
%&  separate & Turns a single  column into multiple columns. & $\tab_{out}.\rows = \tab_{in}.\rows$ \newline  $\tab_{out}.\columns = \tab_{in}.\columns + 1$\\
%\cline{2-4}
    \hline
    \multirow{13}{*}{\begin{turn} {90}\makecell{dplyr}\end{turn}}
&  select & Project a subset of columns in a data frame. & $\tab_{out}.\rows = \tab_{in}.\rows$ \newline $\tab_{out}.\columns < \tab_{in}.\columns$ \\
\cline{2-4}
&  filter & Select a subset of rows in a data frame. &$\tab_{out}.\rows < \tab_{in}.\rows$\newline$\tab_{out}.\columns = \tab_{in}.\columns$\\ 
\cline{2-4}
&  summarise & Summarise multiple values to a single value. &$\tab_{out}.\rows \leq \tab_{in}.\rows$\newline$\tab_{out}.\columns \leq \tab_{in}.\columns + 1$\\ 
\cline{2-4}
&  group\_by & Group a table by one or more variables. &$\tab_{out}.\rows = \tab_{in}.\rows$\newline$\tab_{out}.\columns = \tab_{in}.\columns$\\ 
\cline{2-4}
&  mutate & Add new variables and preserves existing. &$\tab_{out}.\rows = \tab_{in}.\rows$\newline$\tab_{out}.\columns = \tab_{in}.\columns + 1$\\ 
\cline{2-4}
&  inner\_join & Perform inner join on two tables. &${\textsc min}(\tab_{in}^1.\rows,\tab_{in}^2.\rows) \leq \tab_{out}.\rows \leq {\textsc max}(\tab_{in}^1.\rows,\tab_{in}^2.\rows)$\newline$\tab_{out}.\columns \leq \tab_{in}^1.\columns + \tab_{in}^2.\columns - 1$\\ 
\cline{2-4}
\hline
\end{tabular}
}
\caption{Specifications 1 of high-order components}\label{table:spec1}
\vspace{-0.1in}
\end{table*}

On the other hand, as shown in Table~\ref{table:spec2},
\emph{Spec 2} is strictly more precise than \emph{Spec 1}. In addition
to the rows and columns in \emph{Spec 1}, \emph{Spec 2} also uses other
information, such as cardinality and number of groups. For instance, 
$\tab.\groups$ denotes the number of groups in table $\tab$ 
and $\tab.\heads$ denotes the \emph{cardinality of new column names} in table $\tab$ 
with respect to the input table.
Finally, $\tab.\cells$ represents the \emph{cardinality of new values} in table $\tab$
with respect to the input table. Note that the new values includes both new 
column names as well as cell values in $\tab$.
\begin{example}\label{ex:spec2}
Recall the following input table from Example~\ref{ex:long-to-wide}:
%For the input table,
{\small 
 \vspace{0.1in}
\begin{center}
    %\begin{flushleft}{\hspace{1.1in}Raw data:}\end{flushleft}
    %\vspace{0.1in}
\begin{tabular}{c c c c} 
     %Raw data:\\ [0.5ex]
     \hline
     \rowcolor{lightgoldenrodyellow}
     id & year & A & B \\ [0.5ex] 
     \hline\hline
     \rowcolor{babygreen}
     1 & 2007 & 5 & 10\\
     \hline
     \rowcolor{babypurple}
     2 & 2009 & 3 & 50\\
     \hline
     \rowcolor{babygreen}
     1 & 2007 & 5 & 17\\
     \hline
     \rowcolor{babypurple}
     2 & 2009 & 6 & 17\\
     \hline
    \end{tabular}
    \end{center}
     \vspace{0.1in}
    }
For this input table, we use $S_{h1}$ and $S_{c1}$ to represent the set of column names and the set of values, respectively.
Here $S_{h1} = \{\small \texttt{id,year,A,B}\}$ and $S_{c1} = \{\texttt{\small id,year,A,B,1,2,}$\\
$\texttt{\small 3,5,6,10,50,17,2007,2009}\}$. 
Using $S_{h1}$ and $S_{c1}$ we can compute the values of 
$\tab_{in}.\heads$ and $\tab_{in}.\cells$: 
\[
\begin{array}{c}
\tab_{in}.\heads = |S_{h1}-S_{h1}| = 0 \\
\tab_{in}.\cells = |S_{c1}-S_{c1}| = 0
\end{array}
\]
Note that the number of groups in the input table is initialized to 1.

For the output table from Example~\ref{ex:long-to-wide} we can compute the same properties in a similar fashion:
{\small 
 \vspace{0.1in}
    \begin{center}
   % \begin{flushleft}{\hspace{0.4in}Desired data:}\end{flushleft}
   % \vspace{0.1in}
    \begin{tabular}{c c c c c} 
     \hline
     \rowcolor{lightgoldenrodyellow}
     id & A\_2007 & B\_2007 & A\_2009 & B\_2009 \\ [0.5ex] 
     \hline\hline
     \rowcolor{babygreen}
     1 & 5 & 10 & 5 & 17\\
     \hline
     \rowcolor{babypurple}
     2 & 3 & 50 & 6 & 17\\
     \hline
    \end{tabular}
    \end{center}
     \vspace{0.1in}
    }
 
Let $S_{h2}$ and $S_{c2}$ represent the set of column names and the set of values, respectively.
Since $S_{h2} = \{\small \texttt{id,A\_2007,}$ $\texttt{\small B\_2007,A\_2009,B\_2009}\}$ and 
$S_{c2} = \{\small \texttt{id,A\_2007,B\_2007,}$ $\texttt{\small A\_2009,\\B\_2009,1,2,3,5,6,10,50,17}\}$, then we 
can compute $\tab_{out}.\heads$ and $\tab_{out}.\cells$ as follows:

\[
\begin{array}{c}
\tab_{out}.\heads = |S_{h2}-S_{h1}| = 4\\
\tab_{out}.\cells = |S_{c2}-S_{c1}| = 4 
\end{array}
\]

Finally, the number of groups in the output table
is set to a fresh variable $k$ where $k>0$, since we can apply zero or more \texttt{group\_by}
operators before the output table.

Now given the following hypothesis $\hypo$:

\vspace{0.1in}
\Tree [.$?_0^{\texttt{spread}}:\texttt{tbl}$ $?_1:\texttt{tbl}$@$(x_1,$\tab$)$  $?_2:\texttt{cols}$ ]
\vspace{0.1in}
 
if we choose the specification of \texttt{spread} from Table~\ref{table:spec1}, 
the constraint generation function $\Phi(\hypo)$ yields the following Presburger arithmetic formula $\psi$:
\[
\begin{array}{c}
?_0.\rows \leq ?_1.\rows \ \land \  ?_0.\columns \geq ?_1.\columns \ \land \\
?_0.\rows = 2 \ \land \ ?_0.\columns = 5 \ \land \ ?_1.\rows = 4 \ \land \ ?_1.\columns = 4
\end{array}
\]
Since formula $\psi$ is satisfiable, \toolname will continue to explore possible 
completions of hypothesis $\hypo$ even though none of them will lead to a 
correct solution.

On the other hand, if we choose a more precise specification of \texttt{spread} 
presented on Table~\ref{table:spec2}, the deduction system can prune this 
incorrect hypothesis $\hypo$. Here is the new constraint $\psi'$ based on \emph{Spec 2}:
\[
\begin{array}{c}
?_0.\rows \leq ?_1.\rows \ \land \  ?_0.\columns \geq ?_1.\columns \ \land \\
?_0.\rows = 2 \ \land \ ?_0.\columns = 5 \ \land \ ?_1.\rows = 4 \ \land \ ?_1.\columns = 4 \ \land \\
?_0.\groups =  ?_1.\groups\ \land \ ?_0.\cells \leq ?_1.\cells \ \land \\ 
\text{\underline{\boldmath$?_0.\heads \leq ?_1.\cells \ \land \ ?_0.\heads = 4$}} \\ 
\text{\underline{\boldmath$?_1.\cells = 0$}} 
\ \land  \ ?_1.\heads = 0 \ \land \ ?_0.\cells = 4 \ \land \\ ?_1.\groups = 1 \ \land  \ ?_0.\groups = k \ \land k > 1
\end{array}
\]
\end{example}
The above constraint $\psi'$ is unsatisfiable because of the underlined
conjuncts. As a result the deduction will reject hypothesis
 $\hypo$ without completing it.

\begin{table*}
\vspace{-0.1in}
\centering
\newcolumntype{A}{ >{\arraybackslash} m{0.41\textwidth} }
\newcolumntype{B}{ >{\centering\arraybackslash} m{0.085\textwidth} }
\newcolumntype{E}{ >{\centering\arraybackslash} m{0.41\textwidth} }
\newcolumntype{D}{ >{\centering\arraybackslash} m{0.03\textwidth} }
\scalebox{0.9}{
\begin{tabular}{| D | B | A | E |}
   \hline
    Lib & Component & \centering Description & Specification \\
\hline
    \multirow{14}{*}{\begin{turn} {90}\makecell{tidyr}\end{turn}}
&  spread & Spread a key-value pair across multiple columns. &  $\tab_{out}.\groups = \tab_{in}.\groups$ \newline $\tab_{out}.\cells \leq \tab_{in}.\cells$ \newline $\tab_{out}.\heads \leq \tab_{in}.\cells$ \newline $\tab_{out}.\rows \leq \tab_{in}.\rows$ \ ; \ $\tab_{out}.\columns \geq \tab_{in}.\columns$\\
\cline{2-4}
&  gather & Takes multiple columns and collapses into key-value pairs, duplicating all other columns as needed.& 
$\tab_{out}.\groups = \tab_{in}.\groups$ \newline $\tab_{out}.\cells \leq \tab_{in}.\cells + 2$ \newline $\tab_{out}.\heads \leq \tab_{in}.\heads + 2$ \newline $\tab_{out}.\rows \geq \tab_{in}.\rows$ \ ; \ $\tab_{out}.\columns \leq \tab_{in}.\columns$\\
\cline{2-4}
&  separate & Separate one column into multiple columns.& 
$\tab_{out}.\groups = \tab_{in}.\groups$ \newline $\tab_{out}.\cells \geq \tab_{in}.\cells + 2$ \newline $\tab_{out}.\heads \leq \tab_{in}.\heads + 2$ \newline $\tab_{out}.\rows = \tab_{in}.\rows$ \ ; \ $\tab_{out}.\columns = \tab_{in}.\columns + 1$\\
\cline{2-4}
&  unite & Unite multiple columns into one.& 
$\tab_{out}.\groups = \tab_{in}.\groups$ \newline $\tab_{out}.\cells \geq \tab_{in}.\cells + 1$ \newline $\tab_{out}.\heads \leq \tab_{in}.\heads + 1$ \newline $\tab_{out}.\rows = \tab_{in}.\rows$ \ ; \ $\tab_{out}.\columns = \tab_{in}.\columns - 1$\\
\cline{2-4}
%&  separate & Turns a single  column into multiple columns. & $\tab_{out}.\rows = \tab_{in}.\rows$ \newline  $\tab_{out}.\columns = \tab_{in}.\columns + 1$\\
%\cline{2-4}
    \hline
    \multirow{25}{*}{\begin{turn} {90}\makecell{dplyr}\end{turn}}
&  select & Project a subset of columns in a data frame. & 
$\tab_{out}.\groups = \tab_{in}.\groups$ \newline $\tab_{out}.\cells \leq \tab_{in}.\cells$ \newline $\tab_{out}.\heads \leq \tab_{in}.\heads$ \newline $\tab_{out}.\rows = \tab_{in}.\rows$ \ ; \ $\tab_{out}.\columns < \tab_{in}.\columns$\\
\cline{2-4}
&  filter & Select a subset of rows in a data frame. &
$\tab_{out}.\groups = \tab_{in}.\groups$ \newline $\tab_{out}.\cells \leq \tab_{in}.\cells$ \newline $\tab_{out}.\heads = \tab_{in}.\heads$ \newline $\tab_{out}.\rows < \tab_{in}.\rows$ \ ; \ $\tab_{out}.\columns = \tab_{in}.\columns$\\ 
\cline{2-4}
&  summarise & Summarise multiple values to a single value. &
$\tab_{out}.\groups = \tab_{in}.\groups = \tab_{out}.\rows$ \newline $\tab_{out}.\cells \leq \tab_{in}.\cells + \tab_{in}.\groups + 1$ \newline $0 < \tab_{out}.\heads \leq \tab_{in}.\heads + 1$ \newline $\tab_{out}.\rows \leq \tab_{in}.\rows$ \ \ $\tab_{out}.\columns \leq \tab_{in}.\columns + 1$\\ 
\cline{2-4}
&  group\_by & Group a table by one or more variables. &
$\tab_{out}.\groups \ge \tab_{in}.\groups$ \newline $\tab_{out}.\cells = \tab_{in}.\cells$ \newline $\tab_{out}.\heads = \tab_{in}.\heads$ \newline $\tab_{out}.\rows = \tab_{in}.\rows$ \ ; \ $\tab_{out}.\columns = \tab_{in}.\columns$\\ 
\cline{2-4}
&  mutate & Add new variables and preserves existing. &
$\tab_{out}.\groups = \tab_{in}.\groups$ \newline $\tab_{out}.\heads = \tab_{in}.\heads + 1$\newline $\tab_{in}.\cells  < \tab_{out}.\cells \leq \tab_{in}.\cells + \tab_{in}.\rows$ \newline $\tab_{out}.\rows = \tab_{in}.\rows$ \ ; \ $\tab_{out}.\columns = \tab_{in}.\columns + 1$\\ 
\cline{2-4}
&  inner\_join & Perform inner join on two tables. &
$\tab_{out}.\groups = 1 $ \newline $\tab_{out}.\heads \leq (\tab_{in}^1.\heads + \tab_{in}^2.\heads) $\newline $\tab_{out}.\cells \leq (\tab_{in}^1.\cells + \tab_{in}^2.\cells)$ 
\newline ${\textsc min}(\tab_{in}^1.\rows,\tab_{in}^2.\rows) \leq \tab_{out}.\rows \leq {\textsc max}(\tab_{in}^1.\rows,\tab_{in}^2.\rows)$\newline$\tab_{out}.\columns \leq \tab_{in}^1.\columns + \tab_{in}^2.\columns - 1$\\ 
\cline{2-4}
\hline
\end{tabular}
}
\caption{Specifications 2 of high-order components}\label{table:spec2}
\vspace{-0.1in}
\end{table*}

\fi

\end{document}